\documentclass[aps,prb,preprint, nofootinbib, superscriptaddress, longbibliography]{revtex4-2}
\usepackage{geometry}
\usepackage{float}
\usepackage{caption}
\usepackage{amsmath}
\usepackage{amssymb}
\usepackage{braket}
\usepackage{chemformula}
\usepackage{graphicx}
\usepackage{setspace}
\usepackage{esint}
\usepackage{subscript}
\usepackage[lofdepth]{subfig}
\makeatletter

\usepackage{array}
\usepackage[section]{placeins}
\usepackage{xcolor}
\usepackage{soul}
\usepackage{array,booktabs}
\usepackage{relsize}
\usepackage{hyperref}
\usepackage{ulem}
\usepackage{ctable}
\let\sub\textsubscript

\DeclareUnicodeCharacter{2212}{-}

\newcommand{\beginsupplement}{%
        \setcounter{table}{0}
        \renewcommand{\thetable}{S\arabic{table}}%
        \setcounter{figure}{0}
        \renewcommand{\thefigure}{S\arabic{figure}}%
        \setcounter{section}{0}
        \renewcommand{\thesection}{S\arabic{section}}%
        \setcounter{equation}{0}
        \renewcommand{\theequation}{S\arabic{equation}}%
        \setcounter{footnote}{0}        
     }
     
\newif\iffinal

\iffinal
    \newcommand\riley[1]{}
    \newcommand\ramya[1]{}
    \newcommand\jeff[1]{}
    \newcommand\anupam[1]{}
\else
    \newcommand\riley[1]{{\color{blue}[Riley: #1]}}
    \newcommand\ramya[1]{{\color{red}[Ramya: #1]}}
    \newcommand\jeff[1]{{\color{magenta}[Jeff: #1]}}
    \newcommand\anupam[1]{{\color{purple}[AG: #1]}}
\fi

\newcommand{\mrm}{\mathrm}

\begin{document}

\title{Thermal Resistance at a Twist Boundary and Semicoherent Heterointerface}

\author{Ramya Gurunathan}
\affiliation{Department of Materials Science and Engineering, Northwestern University, Evanston IL 60208, USA}
\author{Riley Hanus}
\email{hanusriley@gmail.com}
\affiliation{George W. Woodruff School of Mechanical Engineering, Georgia Institute of Technology, Atlanta, GA 30332, USA}
\author{Samuel Graham}
\affiliation{George W. Woodruff School of Mechanical Engineering, Georgia Institute of Technology, Atlanta, GA 30332, USA}
\author{Anupam Garg}
\affiliation{Department of Physics and Astronomy, Northwestern University, Evanston IL 60208, USA}
\author{G. Jeffrey Snyder}
\email{jeff.snyder@northwestern.edu}
\affiliation{Department of Materials Science and Engineering, Northwestern University, Evanston IL 60208, USA}

\begin{abstract}
Traditional models of interfacial phonon scattering, including the acoustic mismatch model (AMM) and diffuse mismatch model (DMM), take into account the bulk properties of the material surrounding the interface, but not the atomic structure and properties of the interface itself. Here, we derive a theoretical formalism for the phonon scattering at a dislocation grid, or two interpenetrating orthogonal arrays of dislocations, as this is the most stable structure of both the symmetric twist boundary and semicoherent heterointerface. With this approach, we are able to separately examine the contribution to thermal resistance due to the step function change in acoustic properties and due to interfacial dislocation strain fields, which induces diffractive scattering. Both low-angle Si-Si twist boundaries and the Si-Ge heterointerfaces are considered here and compared to previous experimental and simulation results. This work indicates that scattering from misfit dislocation strain fields doubles the thermal boundary resistance of Si-Ge heterointerfaces compared to scattering due to acoustic mismatch alone. Scattering from grain boundary dislocation strain fields is predicted to dominate the thermal boundary resistance of Si-Si twist boundaries. This physical treatment can guide the thermal design of devices by quantifying the relative importance of interfacial strain fields, which can be engineered via fabrication and processing methods, versus acoustic mismatch, which is fixed for a given interface. Additionally, this approach captures experimental and simulation trends such as the dependence of thermal boundary resistance on the grain boundary angle and interfacial strain energy.
\end{abstract}
\maketitle


\section{Introduction}\label{sct:intro}
Given the ubiquity of interfaces and grain boundaries in engineering materials, estimating their influence on thermal transport is essential to the design of devices like integrated circuits and thermoelectrics, particularly in the age of nanostructuring\cite{Kanatzidis2010, Prasher2014, Sood2018}. The current standard models of thermal boundary resistance (or Kapitza resistance, $R\sub{K}$), which are the acoustic mismatch model (AMM) and diffuse mismatch model (DMM), only consider the properties of the media surrounding the interface, but ignore the interfacial defect structure\cite{Swartz1989, Hickman2020, Prasher2001, Varnavides2019, Schelling2004, Zhang2018, Meng2013}. Even the recently introduced strain mismatch model (SMM)\cite{Varnavides2019}, an \textit{ab inito} framework applied to compute the phonon coupling to the long-range dilatation at a Si-Ge heterointerface, neglects the periodic, local strain fields induced by the interfacial defects\cite{Varnavides2019}.
Therefore, trends in thermal resistance with modifications to the local interface structure can not be easily discerned. Molecular dynamics (MD) simulations, however, have revealed an interplay between $R\sub{K}$ and the interface structure and geometry\cite{Ju2013, Schelling2004, Hickman2020}.

While computational techniques such as MD have been useful in probing the atomic-scale structure at interfaces\cite{Ju2013,Schelling2004, Hickman2020, Cahill2003}, the continuum theory-based model presented here provides valuable insights into phonon scattering sources at the nanometer to micron length scale from strain fields at the interface. Therefore, our theoretical approach to thermal boundary resistance enables the multiscale modeling of thermal materials in an integrated computational materials engineering (ICME) framework\cite{Olson2015, Crocombette2009, Kanatzidis2010}. Atomic simulation-based techniques require separate evaluations for every interface structure, whereas this model provides a critical analytical link between thermal boundary resistance, material properties, and grain boundary configurations. As in most work on this problem to date, we treat the scattering in terms of a fixed perturbation. The dynamical degrees of freedom within the interface are ignored, and inelastic scattering in which an interfacial phonon is absorbed or emitted is thus not accounted for. Treating inelastic boundary scattering remains an open problem, although early molecular and lattice dynamics simulations suggest that coupling to localized interfacial modes can influence optical phonon transport at an interface\cite{Gordiz2016, Giri2017}.

Recent experimental work on grain boundaries and heterointerfaces suggests that insights into the role of interfacial dislocation structure on thermal resistance will be highly impactful\cite{Hopkins2011d}. For example, the periodic dislocation structure present at low angle grain boundaries has been associated with significant thermal conductivity reductions and improvements in the thermoelectric performance of well-studied materials such as bismuth antimony telluride\cite{Kim2015, Kim2016}. While several experimental investigations exist for the ensemble average interface scattering in a polycrystal, individual grain boundary types are difficult to study. However, recent $R\sub{K}$ measurements using the 3$\omega$ method on fabricated twist bicrystals of Si\cite{Xu2018} and Al$_2$O$_3$\cite{Tai2013} point to evidence of dislocation strain scattering as a dominant mechanism. For example, the thermal boundary resistance $R\sub{K}$ of these twist boundaries is shown to depend on the grain boundary angle, or equivalently, the dislocation spacing, in addition to the interfacial strain energy. In both the Si and Al$_2$O$_3$ twist boundaries, TEM imaging has been used to verify the presence of dislocation arrays at the interface\cite{Tai2013, Xu2018}. Additionally, heterointerfaces are often intentionally created in thermal materials through a variety of nanostructuring techniques including heterostructures, thin film superlattices, and nanoprecipitate boundaries\cite{Kanatzidis2010, Kim2017}. Thus far, it has been difficult to experimentally determine the effect of misfit dislocations---which can in some cases be controlled through annealing and interlayer thickness---on thermal resistance. Our model, which quantifies the relative importance of interfacial dislocation strain versus acoustic mismatch, and can help answer questions about the degree of disregistry at an interface required to suppress phonon transmission. 

\begin{figure}
    \centering
    \includegraphics[width = 0.8\textwidth]{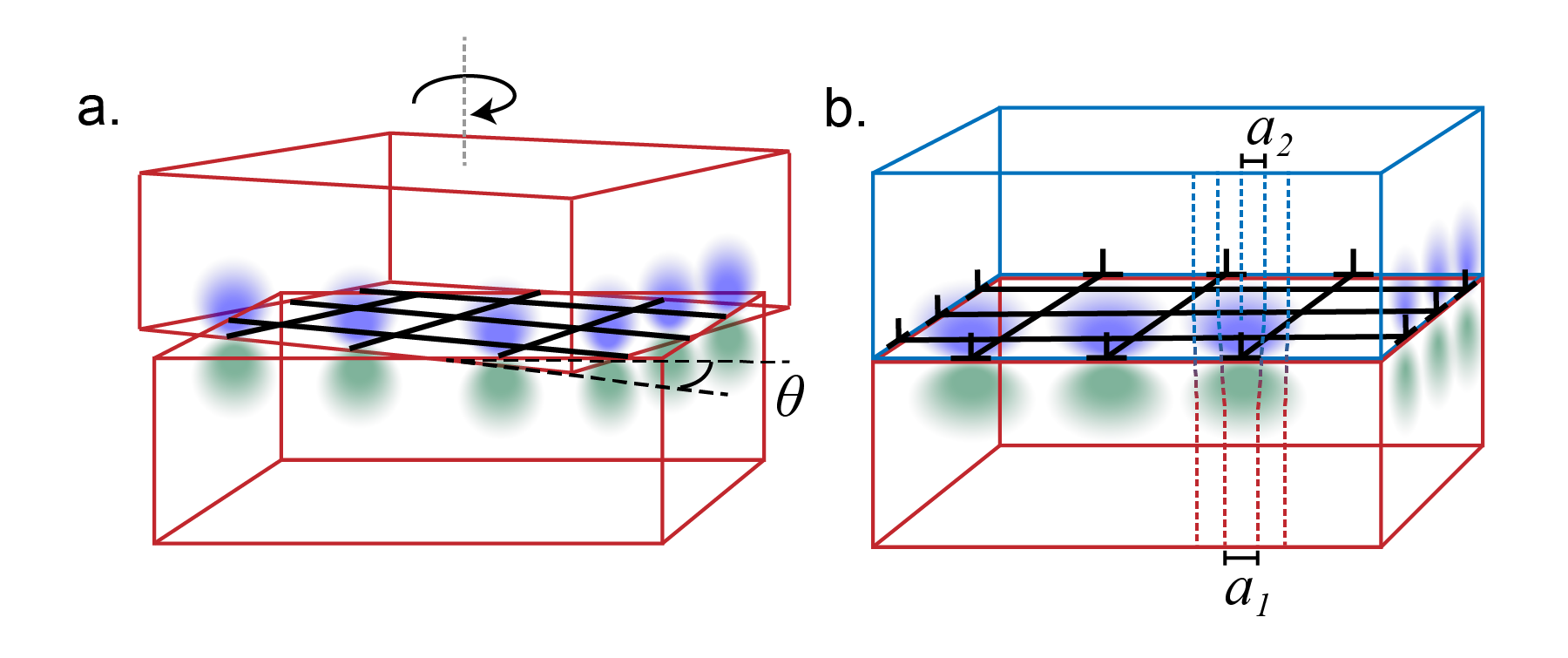}
    \caption{Interfaces described by a grid of linear defects. (a) Schematic of a twist boundary with misorientation angle $\theta$. The black lines indicate screw dislocations and the blue/green shading indicates shear strain. (b) A semicoherent heterointerface between two materials with lattice constants $a_1$ and $a_2$. The black lines indicate edge dislocations and the blue/green shading indicates hydrostatic strain.}
    \label{fig:GB_diagrams}
\end{figure}

In our previous work, a strain scattering theory was used to model the 1D-array of edge dislocations at a symmetric tilt grain boundary, which predicted a frequency($\omega$)-dependent relaxation time, in contrast to standard models\cite{Hanus2018}. The frequency dependence was essential to capture the anomalous low-temperature thermal conductivity trend of $\kappa \propto T^2$ in polycrystalline materials\cite{Hanus2018, Hua2014, Wang2011, Anderson1984}. Here, we extend this framework to interfacial structures in which two interpenetrating arrays of dislocations form a cross-grid. The grid geometry allows us to describe the anharmonic strain scattering of other low-energy grain boundaries common to engineering materials. These include twist boundaries and semicoherent heterointerfaces, which can be decomposed into a grid of screw-type and misfit edge-type dislocations, respectively\cite[pp. 688-700]{CaiNix} (Figure \ref{fig:GB_diagrams}). The dislocation strain scattering mainly affects mid-frequency phonons with wavelengths on the order of the dislocation spacing. High frequency phonons would couple strongly to the atomic inhomogeneities at the dislocation core. We arrive at the interfacial strain scattering potential by superposing single-dislocation-line scattering potentials. This strategy is only applicable for grain boundary angles of less than 15$^{\circ}$ in most materials, because the overlapping dislocation core regions dominate at higher grain boundary angles\cite{Wolf1989}.

\begin{figure}
    \centering
    \includegraphics[width = \textwidth]{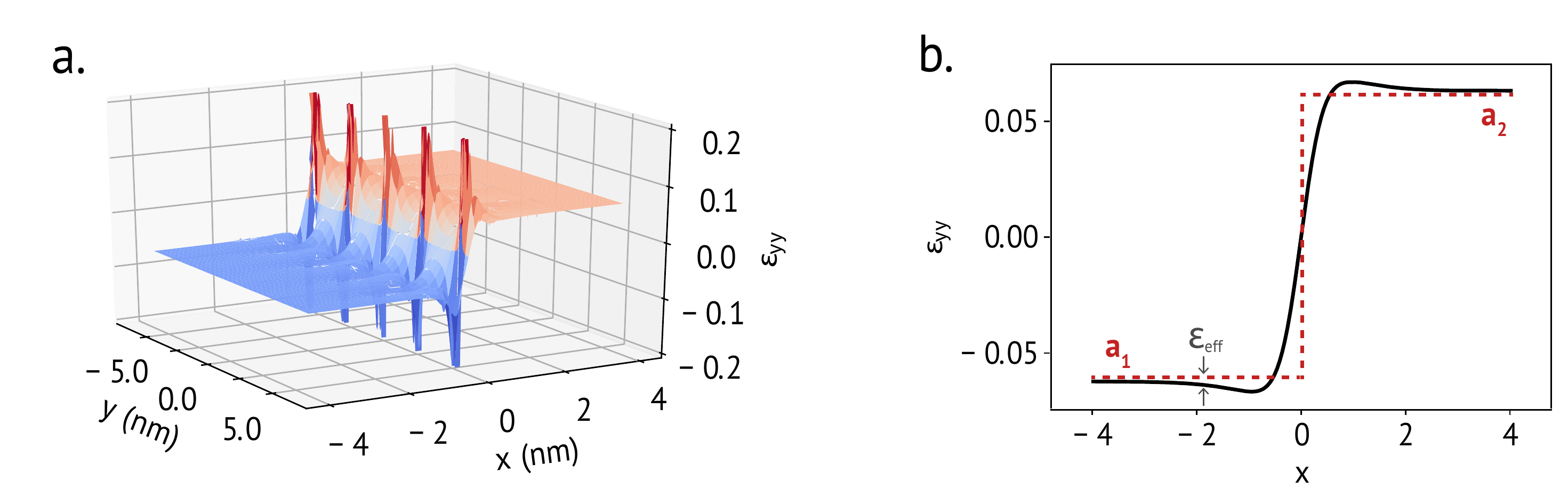}
    \caption{Scattering at a semicoherent heterointerface stems from the periodic strain fields at a misfit dislocation array as well as the step function change in acoustic impedance. (a) Analytic solution for the dilatational strain field component $\epsilon_{yy}$ from an infinite array of misfit dislocations periodically spaced along the $y$-axis\cite[pp. 695-697]{CaiNix}. (b) Cross-section of 3D dilatational strain field, showing an underlying step function (dotted red line). This is indicating a change in lattice parameter ($a$) from material 1 to 2, rather than long-range strain. We subtract off the step function in strain, and instead treat this effect with an acoustic mismatch scattering term. This leaves the physical strain ($\epsilon\sub{eff}$), which we treat with an anharmonic strain scattering potential.}
    \label{fig:misfit}
\end{figure}

There are two main contributions to the scattering potential in this model. The first is the localized strain fields from the dislocation cross-grid, and the second is the mismatch in acoustic properties between sides 1 and 2 of the interface. To illustrate these two contributions, we show in Figure \ref{fig:misfit} the analytically calculated dilatational component $\epsilon_{yy}$ of the strain tensor for a simplified heterointerface in which there is lattice mismatch in one direction only and thus only one array of misfit dislocations. As is evident, the dilatation behaves asymptotically like a step function, but the large nonzero value of $\epsilon_{yy}$ as $|x| \to \infty$ is spurious, since it is being defined with reference to a fictitious average lattice. The actual reference lattice differs on the two sides of the interface. The true or {\it physical strain\/}, $\epsilon\sub{eff}$, must be defined with respect to the true reference lattice, and is obtained by subtracting off the dilatation step function. This strain is much more localized to the vicinity of the interface. As highlighted in Figure \ref{fig:misfit}b, the physical strain scatters via the lattice anharmonicity, while the step function change in lattice parameter and the harmonic properties of the lattice is treated as an acoustic impedance mismatch. One benefit of our approach is the ease of separating the relative scattering contributions of the acoustic mismatch and dislocation strain in each grain boundary type.

In relation to the components of the scattering potential, a novel aspect of our work is the ability to recognize the very different character of phonon scattering with nonzero and zero ${\bf Q}_{\|}$, the (lattice) momentum transfer parallel to the interface. Scattering with ${\bf Q}_{\|} \ne 0$ necessarily arises from the dislocation structure within the interface, and therefore emerges from anharmonic interactions with the resulting strain. By contrast, scattering with ${\bf Q}_{\|} = 0$ washes out the atomistic variations in the interfacial structure, and is dominated by the abrupt change in the {\it harmonic\/} lattice properties in the bulk materials on the two sides of the interface, or acoustic mismatch, analogous to the refraction of light. In the twist boundary case, the acoustic mismatch stems from a rotation of the elastic tensor at the boundary. In the case of a semicoherent heterointerface, it stems from entering a new elastic medium, with an attendant change in the entire elastic tensor.

This separation of the two types of scattering manifests itself in our analysis as follows. In Ref.~\cite{Hanus2018}, it was shown that the periodic array of dislocations behaves like a diffraction grating, whereby the momentum transfer parallel to the dislocation lines vanishes and that in the direction of periodicity is quantized in units of $2\pi/D$ with $D$ being the periodicity or the distance between the dislocation lines. An identical restriction must now apply to each of the two dislocation arrays in the cross grid. Because these arrays are mutually orthogonal, the constraints on momentum transfer imposed by each one separately cannot be satisfied simultaneously, and there is no interference between the scattering from one and the other. The arrays act as essentially independent scatterers. The exception is when the momentum transfer is zero along both the dislocation line directions, i.e., when ${\bf Q}_{\|} = 0$. Such scattering events correspond to either specular reflection or forward scattering, which are precisely what the acoustic mismatch addresses. For these events, therefore, we apply a different scattering potential for acoustic mismatch by modelling it as a step-function change in harmonic phonon properties, specifically the acoustic velocity. While our treatment handles this mismatch scattering within quantum perturbation theory, the magnitude of the transmissivity agrees well with the classical acoustic mismatch model \cite{MonachonTBC2016}. We acknowledge that the limits of perturbation-type methods and their application to phonon-grain boundary or phonon-interface interactions is an open debate. We hope this work demonstrates the utility of this approach.

The plan of the paper is as follows. In Section \ref{subsct:scatt_theory} we present the interface scattering theoretical framework and apply it to a grid of linear defects. We discuss the scattering kinematics, and present the scattering potentials describing both the dislocation strain and the acoustic mismatch.
These potentials are used to calculate a phonon scattering rate $\tau^{-1}$ using Fermi's golden rule. This rate is then used to compute a phonon transmissivity, and in turn, the thermal boundary resistance ($R\sub{K}$). 
Next, we apply the framework to a twist boundary (Si-Si in particular) in Section
\ref{section:twisteky-twisteky},
and to a semicoherent heterointerface (Si-Ge in particular) in Section
\ref{section:hetero}.
For both cases, we compare the $R\sub{K}$ results to experimental measurements and various simulation techniques.
The predicted $R\sub{K}$ values agree with previous computational results from molecular dynamics\cite{Ju2013, Landry2009, Schelling2004}, lattice dynamics\cite{Gordiz2016}, and atomistic Green's function analysis\cite{Zhang2007}, and capture experimentally-observed $R\sub{K}$ trends with grain boundary angle and interfacial energy. Our model also shows that dislocation strain is the majority contribution to $R\sub{K}$ for the Si-Si twist boundary and half of the contribution to $R\sub{K}$ for the Si-Ge heterointerface. Our conclusions are summarized in Section
\ref{section:conclude}. The details of the scattering rate calculation, its relation to the Landauer formalism of phonon transmissivity, and the strain fields of the two types of interface, are given in the Supplementary Notes.

\section{Theoretical Framework}
\label{subsct:scatt_theory}

\noindent The scattering theory employed here uses Fermi's Golden Rule to relate a real-space scattering potential $V(\mathbf{r})$ to a phonon scattering rate $\Gamma(\mathbf{q})$, and follows the notation and methodology of Hanus \textit{et al.}\cite{Hanus2018}. The scattering probability $W_{\mathbf{q}, \mathbf{q'}}$ from phonon mode $\mathbf{q}$ to $\mathbf{q'}$ is defined in terms of the perturbation matrix element $\bra{\mathbf{q}}H'\ket{\mathbf{q'}}$ for the corresponding process while enforcing conservation of energy\cite{Hanus2018}:

\begin{equation}
    W_{\mathbf{q}, \mathbf{q'}} = \frac{2\pi}{\hbar}\left|\bra{\mathbf{q}}H'\ket{\mathbf{q'}}\right|^2 \delta(E_{\mathbf{q'}} - E_{\mathbf{q}}).
    \label{eqn:FGR}
\end{equation}

Restrictions on allowed $\mathbf{q}$ to $\mathbf{q'}$ transitions are discussed in Section \ref{sub:scatt_kin}. The scattering rate or inverse relaxation time $\tau(\mathbf{q})^{-1}$ of a phonon mode $\mathbf{q}$ can then be calculated as the integral of $W_{\mathbf{q}, \mathbf{q'}}$ over all possible final phonon states $\mathbf{q'}$, weighted to suppress non-resistive forward scattering processes. Considering a grain of dimensions\footnote{While the final solution is only dependent on intensive properties, and is therefore independent of the volume or shape of the body, a brick-shaped grain is assumed to simplify the calculation.} $L_x\times L_y\times L_z$, the scattering rate expression simplifies to:

\begin{equation}\label{eqn:inv_tau}
   \tau(\mathbf{q})^{-1} = \Gamma(\mathbf{q}) = \frac{2\pi}{\hbar L_xL_yL_z}\iiint \frac{d^3\mathbf{q'}}{(2\pi)^3}|M(\mathbf{Q})|
^2(1 - \hat{\mathbf{q}}\cdot\hat{\mathbf{q}}')\delta(E_{\mathbf{q'}} - E_{\mathbf{q}}).
\end{equation}

\noindent In the Born approximation, the scattering matrix element $M(\mathbf{Q}) \equiv L_xL_yL_z \bra{\mathbf{q}}H'\ket{\mathbf{q}}$ depends on the scattering vector $\mathbf{Q} = \mathbf{q'} - \mathbf{q}$.
The scattering matrix element can, in turn, be written in terms of a real space scattering potential
$V(\mathbf{r})$, as
\begin{equation}
\label{eqn:FT_matrix_elem}
    M(\mathbf{Q}) = \iiint d^3r\, V(\mathbf{r})e^{i\mathbf{Q}\cdot\mathbf{r}}.
\end{equation}
\noindent We take this grain to contain a single low-angle grain boundary or semicoherent interface planning the entire yz plane. As explained in Section \ref{sct:intro}, depending on whether ${\bf Q}_{\|}$ is nonzero or not, we will model $V(\mathbf{r})$ as the perturbation due to either the strain fields of the periodic dislocation array, or the acoustic mismatch produced by the step function change in elastic properties (see Section \ref{sub:scatt_pot}).

From the $\tau$ derived in Equation \ref{eqn:inv_tau}, we can then compute the phonon transmissivity $\alpha_{12}$ from side 1 to side 2 of the interface. This is done by relating the Landauer and interfacial scattering approaches for computing interfacial thermal resistance\cite{DamesChen2004}. The transmissivity $\alpha_{12}$ can be expressed in terms of the relaxation time $\tau$, side 1 group velocity $v\sub{g1}$, and the distance between interfaces or grain size $L_x$ as:
\begin{equation}
    \alpha_{12} = \frac{v\sub{g1} \tau}{\frac{3}{4}L_x + v\sub{g1}\tau}.
    \label{eq:alpha12_to_tau}
\end{equation}
In Section \ref{sec:Landauer-pert}, we provide a derivation of this equation along with an analysis showing that our perturbation treatment for acoustic mismatch transmissivity agrees within 5\% of the classical acoustic mismatch model (given in Ref. \cite{Swartz1989} and here in Eq. \ref{eq:t_AMM}), even when the relative change in phonon velocity across the interface $\Delta v/v$ is as much as 50\%. 

The end product of our calculation is the thermal boundary conductance (inversely, the thermal boundary or Kapitza resistance, $R\sub{K}$). This is computed from the perspective of either side of the interface using the transmissivity as an input\cite{MonachonTBC2016}. We integrate up to the maximum phonon frequency $\omega\sub{m}$, the product of the side 1 spectral heat capacity $C_1(\omega)$ and group velocity $v\sub{g1}(\omega)$, as well as the transmissivity at the interface from side 1 to side 2, $\alpha_{12}(\omega)$, and the reverse direction, $\alpha_{21}(\omega)$\cite{MonachonTBC2016, DamesChen2004, Chen1998, Chen2013}: 

\begin{equation}\label{eqn:Rk}
1/R\sub{K} = \frac{1}{4}\int_0^{\omega\sub{m}} C_1(\omega) v\sub{g1}(\omega) \left( \frac{\alpha_{12}(\omega)}{1 - \overline{\alpha}(\omega)} \right) d\omega.
\end{equation}
Here, $\overline{\alpha} (\omega) = (\alpha_{12}(\omega) + \alpha_{21}(\omega))/2$. This treatment of the transmissivity factor arises by considering the local equilibrium temperature for incident and outgoing phonons \cite{MonachonTBC2016, Chen1998, Chen2013}. It resolves the Kaptiza paradox where, if the factor in the parenthesis is replaced by $\alpha_{12}(\omega)$, a system with a transmissivity of $\alpha_{12}=1$ would have a non-zero resistance, which is unphysical. The modification of the parenthetical factor is important for interfaces with high transmissivities such as twist boundaries. The transmissivity and phonon velocities are often temperature independent, and so the temperature dependence of $R\sub{K}$ enters solely through the spectral heat capacity\cite{MonachonTBC2016}.

To summarize, a real-space scattering potential $V(\mathbf{r})$ is defined from the defect perturbation and used to compute the perturbation matrix element $M(\mathbf{Q})$. The scattering rate $\tau^{-1}$ is then calculated using Fermi's Golden Rule. From the relaxation time $\tau$, additional transport properties including the phonon transmissivity $\alpha$ and thermal boundary resistance $R\sub{K}$ are computed.

We would like to note that our implementation of the model is not polarization specific. Therefore, we do not evaluate the complex mode conversions at the boundary of two elastic solids. In this aspect, our implementation is similar to that of analytic models such as the AMM and DMM\cite{MonachonTBC2016}. The phonons in this model are described by a Born-von Karman dispersion parameterized using the mode-averaged speeds of sound listed in Table \ref{tab:param_vals}.

The following subsections provide additional details about the scattering kinematics of this problem, as well as the strain and acoustic mismatch scattering potentials present at grain boundaries and heterointerfaces.

\subsection{Scattering Kinematics}\label{sub:scatt_kin}

\begin{figure}
    \centering
    \includegraphics[width = 0.7\textwidth]{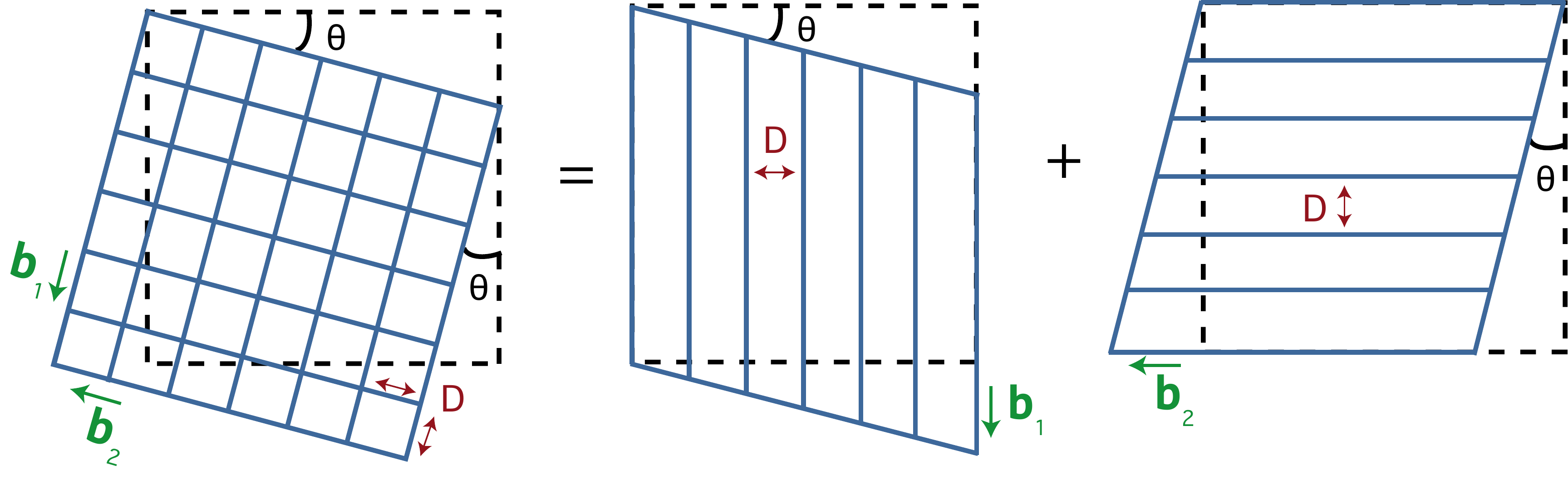}
    \caption{Schematic of two orthogonal screw dislocation arrays, with Burgers vectors $\mathbf{b_1}$ and $\mathbf{b_2}$, respectively, producing a twist misorientation ($\theta$) at an interface.}
    \label{fig:twist_schem}
\end{figure}

\begin{figure}
    \centering
    \includegraphics[width = \textwidth]{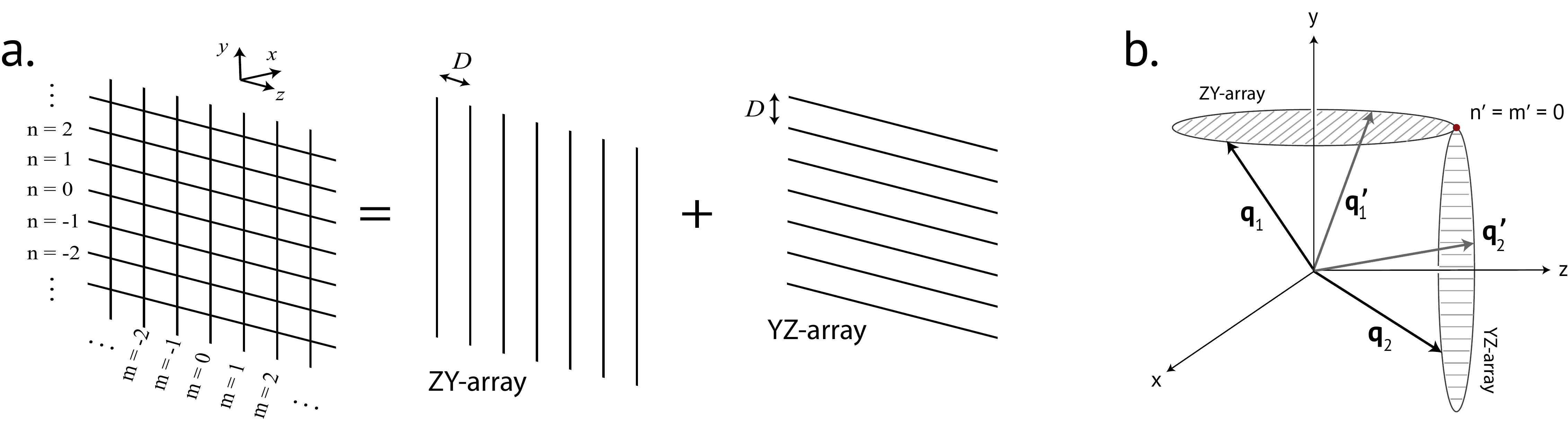}
    \caption{(a) Diagram of orthogonal YZ and ZY arrays in dislocation cross-grid. In this case, equal $D$ spacing is assumed for both. (b) Phase space diagram portraying the independent scattering of the YZ and ZY dislocation array, which overlap only at the ${\bf Q}_{\|} = 0$ ($n'=m'=0$) condition.}
    \label{fig:twist_grid}
\end{figure}

As mentioned, several grain boundary geometries are composed of two dislocation arrays forming a cross-grid. In the twist boundary case, for example, two sets of screw dislocation arrays each shear the crystal to induce a full rotation (see Figure \ref{fig:twist_schem})\cite{GunterGottstein2009}. We adopt the configuration in Figure \ref{fig:twist_grid}a, with the $x$-direction normal to the interface, and two orthogonal dislocation arrays with dislocation lines in the $y$ and $z$ direction. We will refer to the first dislocation array as the YZ array, where the first label (y) indicates the direction of periodicity, and the second label (z) indicates the direction of the dislocation line (see Figure \ref{fig:twist_grid}a). The second array is likewise called the ZY array. The scattering potential of the cross-grid is given by summing over the single-dislocation-line scattering potentials ($V_1$; see Section \ref{subsub:disl_strain}) for each array and then combining both,

\begin{equation}\label{eqn:Vgrid_sum}
    V(\mathbf{r})=\sum_{n=-\infty}^{\infty}V_{1}(x,y-nD)+\sum_{m=-\infty}^{\infty}V_{1}(x,z-mD).
\end{equation}

\noindent Here, $n$ and $m$ can assume all integer values from $-\infty$ to $+\infty$. The infinite sums over $n$ and $m$ can be obtained analytically for both the twist and heterointerface cases \cite{VanDerMerwe1950, CaiNix} and are shown in Section \ref{suppsct:twist}.

For simplicity, we'll focus on the contribution of the YZ array in the following steps, but analogous expressions can be written for the ZY array, by instead enforcing periodicity in the $z$-direction. The Fourier transform of this sum of dislocation scattering potentials is:

\begin{align}
\label{eqn:Vtilde_sum}
\widetilde{V}\sub{YZ}(Q_x, Q_y) & =\iint dx\,dy\sum_{n=-\infty}^{\infty}V_{1}(x,y-nD)e^{-i(Q_{x}x+Q_{y}y)}\nonumber \\
 & =\sum_{n=-\infty}^{\infty}e^{-iQ_{y}nD}\widetilde{V}_{1}(Q_{x},Q_{y}).
\end{align}.

\noindent We show the Fourier transform of the scattering potential ($\widetilde{V}_1$) as a function of only $Q_x$ and $Q_y$, because the scattering vector along the line of the dislocation ($Q_z$) is necessarily 0\cite{Carruthers1959}. By the Poisson summation formula (see Equation \ref{eq:PoissonSum}), this simplifies to
\begin{equation}
\widetilde{V}\sub{YZ}=\frac{2\pi}{D}\sum_{n'=-\infty}^{\infty}\delta(Q_{y}- Q_{n'})\widetilde{V}_{1}(Q_{x},Q_{n'}),\quad \mrm{with} \quad \Bigl(Q_{n'}=\frac{2\pi n'}{D}\Bigr).
\label{eq:V-tilde-Qn}
\end{equation}

\noindent As noted in Hanus \textit{et al.}\cite{Hanus2018}, this equation shows that phonon diffraction peak conditions will occur whenever the magnitude of the scattering wavevector component $Q_y$ equals $2\pi n'/D$ in an infinite interface\cite{Carruthers1959, Carruthers1961, Omini2000}. Equation \ref{eqn:FT_matrix_elem} can then be used to calculate the perturbation matrix element, resulting in,

\begin{equation}
    \label{eqn:matrix_elem_grid}
    M(\mathbf{Q}) = 2\pi\delta(Q_{z})\widetilde{V}\sub{YZ}(Q_{x},Q_{y})+2\pi\delta(Q_{y})\widetilde{V}\sub{ZY}(Q_{x},Q_{z})\,.
\end{equation}

As enforced by the $\delta$-functions, the scattering due to the YZ array is only non-zero when $Q_z = 0$, while scattering due to the ZY array is only non-zero when $Q_y = 0$. As a result, except when $Q_y = Q_z = 0$, the two dislocation arrays scatter independently (see Figure \ref{fig:twist_grid}b). The Van der Merwe method for calculating interfacial strain energies\cite{VanDerMerwe1950} makes a similar assertion, namely that the energy of both arrays can be reasonably computed separately and then superposed. Specifically analyzing the $Q_y = Q_z = 0$ condition reveals that this scenario must represent either a non-resistive forward scattering case or a mirror-like reflection. The underlying, periodic structure of the interface is washed out at this long-wavelength limit. We treat this scattering separately in terms of the acoustic impedance mismatch (see Section \ref{sec: rotation}). The total scattering rate $\Gamma$ is then the sum of the rates due to the YZ and ZY array, as well as that due to acoustic mismatch (AM) (see Equation \ref{rot_term}):
\begin{equation}\label{eq:gamma_grid_total}
    \Gamma\sub{tot} = \Gamma\sub{YZ} +  \Gamma\sub{ZY} + \Gamma\sub{AM}.
\end{equation}

\noindent Here, $\Gamma\sub{YZ}$ entails only $|\widetilde{V}\sub{YZ}|^{2}$, $\Gamma\sub{ZY}$ entails only $|\widetilde{V}\sub{ZY}|^{2}$, and $\Gamma\sub{AM}$ entails only $|\widetilde{V}\sub{AM}|^{2}$ (see Equations \ref{eqn:V_tilde_R} and \ref{rot_term}). To avoid misunderstanding, we note that the resemblance of Equation \ref{eq:gamma_grid_total} to Matthiessen's Rule is superficial. The $\Gamma\sub{AM}$ component does not represent a separate scattering channel but rather a completely independent kind of interface scattering, which is, in addition, activated at a different frequency regime.

Specifically, the scattering rate $\Gamma\sub{YZ}$ due to periodic strain from the YZ array is:

\begin{equation}
\Gamma\sub{YZ}(\mathbf{q})=\frac{n\sub{b}}{\hbar^{2}D^{2}}\sum_{n'=-\infty}^{\infty}\iiint d^{3}\mathbf{q}^{\prime}\,\delta(Q_{z})\delta(\omega_{\mathbf{q}}-\omega_{\mathbf{q^{\prime}}})\left|\widetilde{V}\sub{YZ}(Q_{x},Q_{y})\right|^{2}(1-\hat{\mathbf{q}}\cdot\hat{\mathbf{q}}').\label{eq:Gamma-1_simplified}
\end{equation}

\noindent Here, $n\sub{b}$ is equal to $1/L_x$, and represents the linear density of boundaries in the material. The result for $\Gamma\sub{ZY}$ is similar.

With the scattering constraints imposed by defect geometry handled, the final step is to define scattering potentials from the interface properties. In the next section, we derive an anharmonic scattering potential from the interfacial dislocation strain fields, as well as a scattering potential from acoustic mismatch, which couples to phonons via harmonic elastic constants.

\subsection{Scattering Potentials}\label{sub:scatt_pot}
\subsubsection{Dislocation Strain}\label{subsub:disl_strain}
The real-space strain scattering potential or lattice energy perturbation is directly related to the induced internal strain $\epsilon(\mathbf{r})$ at the interface via an anharmonic coefficient, which in this case is the Gr\"uneisen parameter ($\gamma = (1/\omega)d\omega/d\epsilon$). A single Gr\"uneisen parameter approximation is made wherein $\gamma$ is frequency and mode-independent, so that the change in phonon frequency due to internal strain is $\omega\gamma\epsilon(\mathbf{r})$. This approximation may lead to an underestimation of the phonon scattering, but trends with misorientation and comparisons of grain boundary geometry should still hold\cite{Hanus2018, Klemens1955}. The scattering potential due to the strain from a single interfacial dislocation, $\epsilon_1(\mathbf{r})$, is:

\begin{equation}\label{eqn:single_disl_V}
V_1(\mathbf{r}) = \hbar \omega \gamma \epsilon_1(\mathbf{r}).
\end{equation}

We use the dislocation strain fields from continuum elasticity theory as given, for example, by Hirth and Loethe\cite[pp. 60, 76]{Loethe1982}. As discussed in the previous section, the sum over single-line-dislocation potentials is facilitated in Fourier space, so all we require is the Fourier transformed strain fields for a single screw and misfit-edge dislocation (similar to methods proposed in Refs. \cite{Varnavides2019, Hanus2018}). These are provided in Sections \ref{subsct:twisteky_strain} and \ref{subsct:hetint_details}, respectively.


\subsubsection{Acoustic Mismatch \label{sec: rotation}}

In the case of the semicoherent heterointerface, the acoustic impedance mismatch stems from the change in material, and resulting change in elastic tensor, across the interface. In the twist boundary case, the long-range rotational deformation induces acoustic mismatch through the anisotropy of the acoustic properties. The rotation at a twist boundary is described by a single misorientation angle $\theta$ (see Figure \ref{fig:twist_schem}), which in the Read-Shockley model, relates to the magnitude of the Burger's vector ($b$) and the dislocation spacing ($D$) as: $2\mathrm{tan}(\theta/2) = b/D$\cite[p. 688]{CaiNix}. For a fixed phonon angle of incidence, the crystal rotation can be interpreted as a change in the acoustic impedance stemming from the rotation of the stiffness tensor. For both grain boundaries and heterointerfaces, an acoustic mismatch scattering potential can be determined, which is grounded in the same physics as the classical acoustic mismatch model (AMM)\cite{Schelling2004}. 

We define the scattering potential as the change in energy of a phonon as it traverses an interface, which can be expressed in terms of the change in phonon phase velocity $\Delta v\sub{p}$ and incident phonon wavevector magnitude $q$,

\begin{equation}
    V_\mathrm{AM}(\mathbf{r})= \hbar \Delta \omega (\mathbf{r}) = \hbar \Delta v_\mathrm{p}(\mathbf{r}) q.
    \label{eq:V_AM}
\end{equation}

The spatial dependence of $\Delta v_\mathrm{p}(\mathbf{r})$ is taken as $\Delta v \, \Theta(x)$ where $\Delta v = v_2 - v_1$ is the magnitude of the phonon velocity change from side 1 to side 2, and $\Theta(x)$ is the Heaviside step function. Since its Fourier transform is $\widetilde{\Theta}(Q_x)=i/(Q_x)$,
\begin{equation}
\widetilde{V}\sub{AM} = \hbar \, \Delta v \, q \, \widetilde{\Theta}(Q_x) = \hbar \Delta v \frac{iq}{ Q_x}.
\label{eqn:V_tilde_R}
\end{equation}
The magnitude of the velocity change $\Delta v$ depends on the phonon angle of incidence and the degree of misorientation at a grain boundary or homointerface, and, at a heterointerface, the additional change in elastic tensor. We use the solver provided by Jaeken \textit{et al.} \cite{Jaeken2016} to solve the Christoffel equation (which is essentially the classical limit of the lattice dynamical matrix diagonalization), and compute the direction-dependent group and phase velocities of the acoustic phonons directly from the stiffness matrix (Supplementary Section \ref{supp:christ_eq}). From these direction-dependent velocities, we can calculate $\Delta v$ for an incoming phonon and capture the acoustic mismatch due to any grain boundary misorientation or change in elastic coefficients, regardless of crystal symmetry. Our implementation lies within the continuum, long-wavelength limit, and so $V\sub{AM}$ computed using the magnitude of either the group or the phase velocity yields the same result given that the perturbation is set only by the change in phonon frequency. The acoustic mismatch constitutes planar defect scattering, and as mentioned previously, will produce a specular reflection. Since forward scattering does not contribute to the scattering rate, $Q_x$ in Equation \ref{eqn:V_tilde_R} will simplify to $2q_x$. 

Our treatment agrees with the conceptual conclusions from the work of Brown\cite{Brown1983}, which suggests that rotations of the crystal scatter phonons via harmonic elastic constants while strain scatters via third order elastic constants\cite{Brown1983}. In Section \ref{sec:Landauer-pert} Figure \ref{fig:Mattheissen-vs-Landauer}b, we compare this perturbation theory treatment of acoustic mismatch scattering to the classical AMM result and show that they agree within 5\% for velocity mismatches typical of solid-solid interfaces. To avoid issues with the change of reference frame, especially when handling the scattering effects due to the long-range deformation, we define the wavevector directions and the lattice perturbations with reference to a virtual average lattice and assure that the scattering potential is symmetric at the boundary\cite{Carrete2017}. 

Finally, the square of the Fourier space scattering potential must be taken in the calculation of the matrix element ($M$). In the work of Brown\cite{Brown1983}, it was shown that symmetry constraints in the cubic crystal enforce that strain and rotation contribute independently to the scattering potential. We reach the same conclusion in our work by noting the distinct scattering physics of strain (diffractive scattering) and rotation (specular reflection). As a result, there are no non-zero cross-terms when we take the square and the full, squared scattering potential can be written as,

\begin{equation}\label{rot_term}
    |\widetilde{V}(\mathbf{Q})|^2 = |\hbar \omega \gamma \widetilde{\epsilon}(\mathbf{Q})|^2 + |\hbar \Delta v \widetilde{\Theta}(x)q|^2.
\end{equation}

\begin{figure}
    \centering
    \includegraphics[width = 0.8\textwidth]{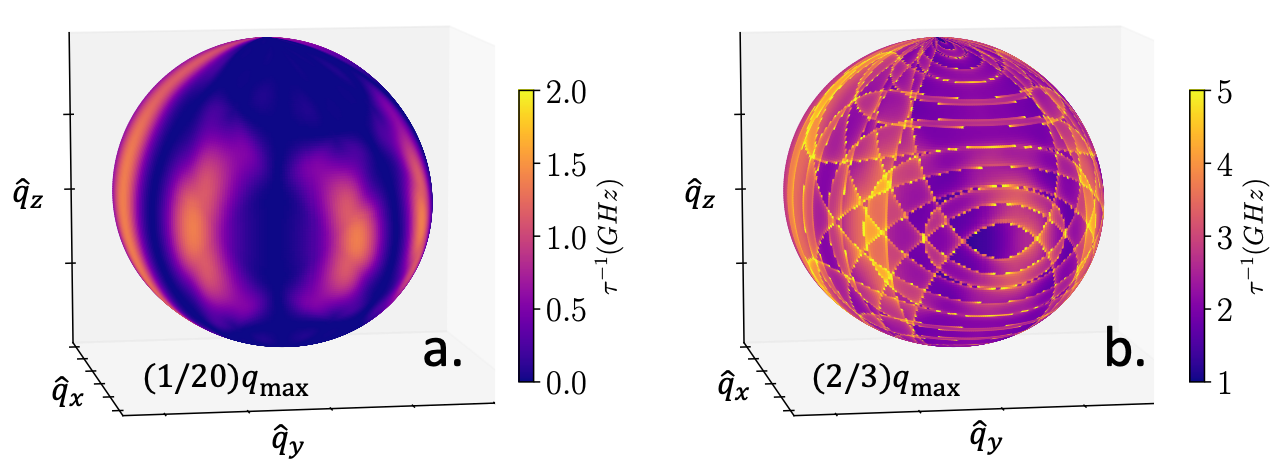}
    \caption{Three-dimensional polar plots of the scattering rate $\tau^{-1}(\mathbf{q})$ (in GHz) versus incident angle ($\theta_i$,$\phi_i$) of an incoming phonon, holding phonon frequency constant. The results shown correspond to a twist boundary with $\theta$ = 5$^\circ$ at the (a) long wavelength limit ($q = q\sub{max} / 20$), where acoustic mismatch scattering dominates and the (b) short wavelength limit ($q = (2/3)q\sub{max}$), where the periodic strain field scattering effect is picked up.}
    \label{fig:tau_directional}
\end{figure}

Figure \ref{fig:tau_directional} shows the phonon scattering rate plotted versus the incident angle of the incoming phonon for a twist boundary. It provides a visual representation of the rotation versus strain scattering effects. At high frequency, the diffraction effects stemming from the periodic dislocation array are visible as patterns in the directional plot of scattering rate. Whereas at low frequency, there are more isolated and broad scattering ``hot spots" corresponding to ranges in the phonon angle of incidence which undergo large scattering due to an acoustic impedance mismatch\cite{Meng2013}. 

\section{Twist Boundary Scattering}
\label{section:twisteky-twisteky}

In this section, we aim to construct a model that explicitly considers the strain effects of the interfacial screw dislocations present at low-angle twist boundaries, while also treating the rotational deformation. 

\subsection{Twist Boundary Strain Fields}\label{subsct:twisteky_strain}

The strain scattering potential of a screw dislocation grid can be described using Equations \ref{eqn:Vgrid_sum} and \ref{eqn:single_disl_V}. We will maintain the geometry of the previous section with the $x$-direction normal to the interface and dislocation arrays with sense vectors oriented along the $y$ and $z$ directions. The strain state of a twist boundary is pure shear, such that all components $\epsilon_{ii}$ are 0. Only two independent components of the strain tensor are non-zero for each dislocation array. These are given in real space in Section \ref{suppsct:twist}. Table \ref{tbl:strain_fields} lists the strain component Fourier transforms for a constituent screw dislocation in either the YZ or ZY array:

\begin{table}[h]
\caption{Twist Boundary Fourier Strain Field Components}
\label{tbl:strain_fields}
\centering
\large
\newcolumntype{C}{>{\centering\arraybackslash} m{6cm} } 
\begin{tabular}{|C|C|}
\hline
   YZ array  & ZY array \\
\hline
\vspace{5pt}
 $\mathlarger{\widetilde{\epsilon}_{13}=\frac{ibQ_{y}}{2(Q_{x}^{2}+Q_{y}^{2})}}$ &
 \vspace{5pt}
 $\mathlarger{\widetilde{\epsilon}_{12}=-\frac{ibQ_{z}}{2(Q_{x}^{2}+Q_{z}^{2})}}$\\[20pt]
 \hline 
 \vspace{5pt}
 $\mathlarger{\widetilde{\epsilon}_{23}=-\frac{ibQ_{x}}{2(Q_{x}^{2}+Q_{y}^{2})}}$ &
 \vspace{5pt}
 $\mathlarger{\widetilde{\epsilon}_{23}=\frac{ibQ_{x}}{2(Q_{x}^{2}+Q_{z}^{2})}}$\\[20pt]
\hline
\end{tabular}
\end{table}

It is a textbook result that while a single array of screw dislocations produces a long-range shear stress, the two periodic screw dislocation arrays in a twist boundary cancel each other's long range stress field\cite[pp. 699-700]{CaiNix}. This cancellation can be seen in the long-wavelength limit of the Fourier-transformed strain fields, as $Q_y$ approaches 0 in the YZ array and $Q_z$ approaches 0 in the ZY array. The components $\widetilde{\epsilon}_{13,}$, $\widetilde{\epsilon}_{12, }$, and the sum of the two $\widetilde{\epsilon}_{23}$ parts all vanish. The cancellation is additionally apparent in the $|x| \to \infty$ limit of the real-space analytic solutions for the strain at infinite screw dislocation arrays, as shown in Supplementary Section \ref{suppsct:twist} \cite{CaiNix}. We can manually enforce this by omitting the $n'=0$ term in the calculation of $\widetilde{V}\sub{YZ}$ or $\widetilde{V}\sub{ZY}$ (Equation \ref{eq:V-tilde-Qn}) and instead treating the corresponding scattering term in the long wavelength limit with the acoustic mismatch term $\widetilde{V}\sub{AM}$ as shown in Equation \ref{eqn:V_tilde_R}.

\subsection{Twist Boundary Results and Discussion}
The results discussed here describe \ch{Si}-\ch{Si} symmetric twist boundaries with various twist angles $\theta$, and were calculated using the \ch{Si} parameters in Table \ref{tab:param_vals}.

\begin{figure}
    \centering
    \includegraphics[width = \textwidth]{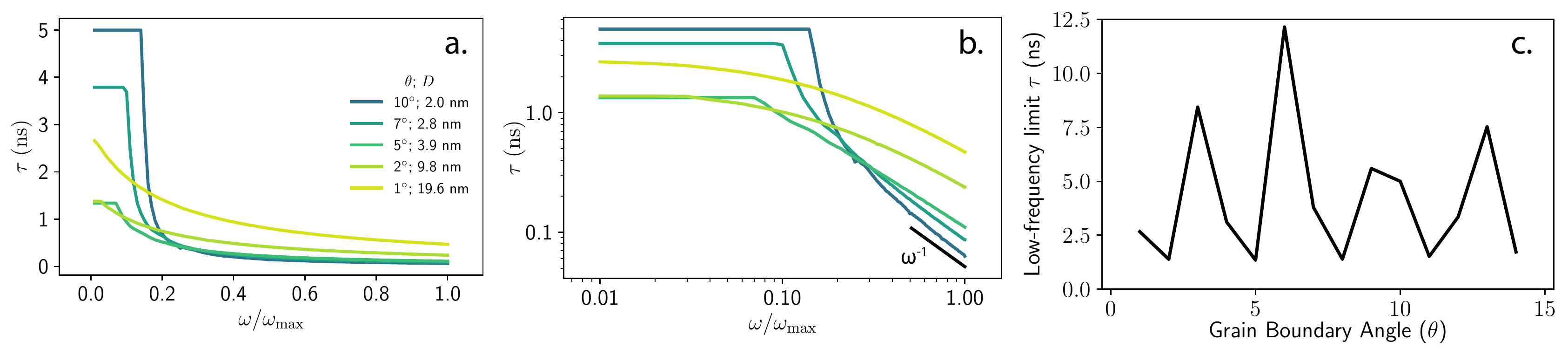}
    \caption{Si-Si twist boundary scattering relaxation times (a) Spectral phonon relaxation times for a Si-Si twist interface at various grain boundary angles (b) The log-log plot of this relaxation time shows a power law crossover from $\omega$-independent to $\sim\!\omega^{-1}$. (c) The long-wavelength limit of the relaxation time is plotted against grain boundary angle, revealing a periodic variation.}
    \label{fig:tau_spectral}
\end{figure}

As previously noted in the case of the tilt boundary, the twist boundary exhibits a cross-over in the frequency-dependence of the relaxation time ($\tau$)\cite{Hanus2018}. Long-wavelength phonons view the boundary as a planar defect defined by the rotational deformation, leading to the expected frequency-independent scattering. The relaxation time at this long wavelength, or low frequency, limit is plotted versus grain boundary angle in Figure \ref{fig:tau_spectral}c and is seen to vary periodically with angle $\theta$. This periodic relationship has been predicted previously and is a result of the symmetry of the \ch{Si} acoustic properties\cite{Brown1983}. Short wavelength phonons, however, interact with the underlying periodic strain from the dislocation grid and pick up a phonon frequency dependence approaching $\tau \propto \omega^{-1}$ (Figure \ref{fig:tau_spectral}b). As a result, the overall thermal boundary resistance increases linearly with grain boundary angle because of the increasing strain scattering effects. 

\begin{figure}
    \centering
    \includegraphics[width = \textwidth]{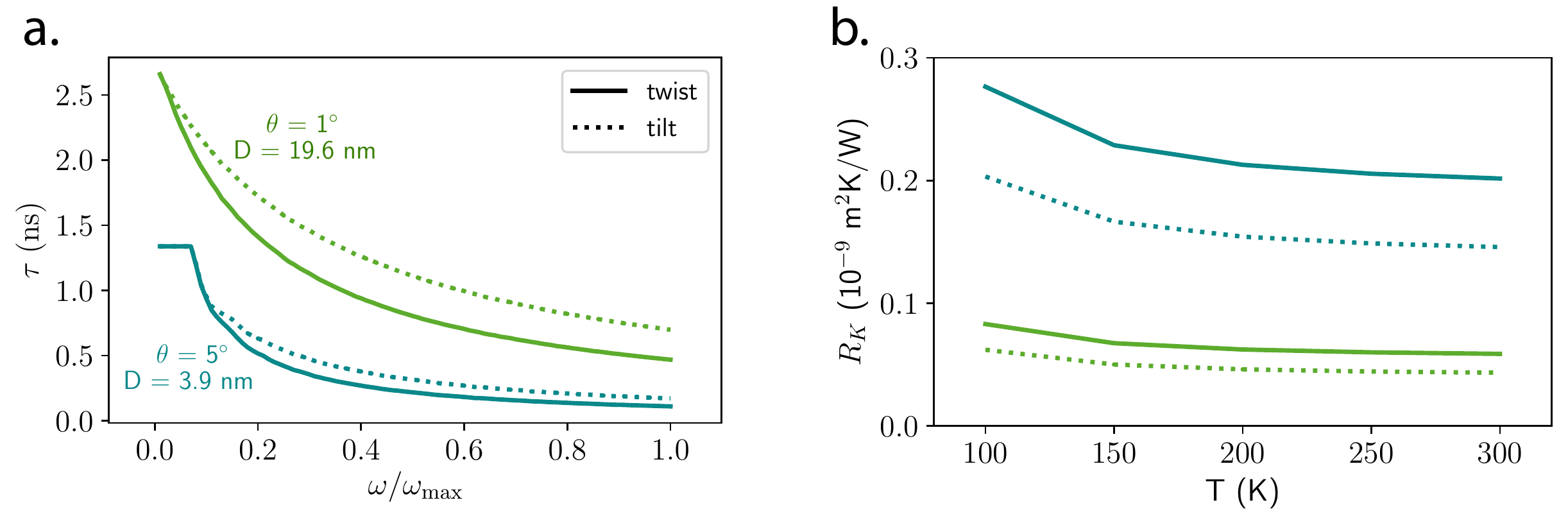}
    \caption{The comparison of twist and tilt boundaries show that: (a) The spectral relaxation time ($\tau$) converges at the long-wavelength limit, but decreases faster in the frequency-dependent regime for the twist boundary. (b) The twist boundary, therefore, is predicted to have about 1.3$\times$ the thermal boundary resistance ($R\sub{K}$).}
    \label{fig:tilt_twist}
\end{figure}

We can also compare these results to the symmetric tilt boundary scattering case (Figure \ref{fig:tilt_twist}). In both grain boundary types, the rotational scattering is calculated from Equation \ref{eqn:V_tilde_R} using the grain boundary angle $\theta$ to determine the phonon velocity change at the interface. It should be kept in mind, however, that in the tilt case, the rotation is perpendicular to the plane of the interface.
The spectral $\tau$ in both cases is approximately equal at the long-wavelength limit as a result of the cubic symmetry of the \ch{Si} stiffness matrix and acoustic velocities (Supplementary Figure \ref{fig:slowness_plot}). However, for the twist boundary, the relaxation time decreases more rapidly with frequency in the dislocation scattering regime (Figure \ref{fig:tilt_twist}a). In the work of Van der Merwe\cite{VanDerMerwe1950}, a linear elasticity model for interfacial stresses and energies is applied to a generic material with cubic or tetragonal symmetry, and shows that for the same misorientation angle $\theta$, twist boundaries exhibit slightly higher strain energy than tilt boundaries. The higher strain energy of the twist boundary can explain the reduced relaxation times at high phonon frequency, which leads to about 1.3$\times$ the thermal boundary resistance of the tilt boundary (Figure \ref{fig:tilt_twist}b).

The $R\sub{K}$ from the \ch{Si}-\ch{Si} twist boundary model (Equation \ref{eqn:Rk}) is close to, although consistently lower than, previously reported molecular dynamics simulation results (see Table \ref{tab:twist_comp}). We also compare model predictions against $R\sub{K}$ measurements of \ch{Si}-\ch{Si} twist boundaries using the 3$\omega$ method, an AC technique suited for thermal conductivity measurements of films, reported in Xu \textit{et al}\cite{Xu2018}. In both cases, the magnitude of the thermal resistance depends on twist angle, and the $R\sub{K}$ ratio between the 6.9$^\circ$ and 3.4$^\circ$ twist boundaries is similar. However, the measured thermal resistance is more than an order of magnitude larger than the model predictions. The model assumes a clean interface, while the interface in the physical material serves as a sink for additional defects and may contain roughness or oxidation effects\cite{Yu2020, Khafizov2014}. In this particular experiment, a Si thin film was bonded to a Si substrate at varying twist angles. TEM images revealed a nanometer-thick disordered region at the boundary, which contributes additional thermal resistance\cite{Xu2018, Shin2014, Li2019}. A detailed modeling of these contributions is necessary to understand the experimental results.

A benefit of our approach is the ability to differentiate between the scattering contributions of the rotational deformation and the dislocation strain. The percentage contribution of the acoustic mismatch effect to $R\sub{K}$ is only about 4-5\% for most symmetric twist boundaries, and dislocation strain accounts for the rest of the scattering. This breakdown illustrates the significant role of the interfacial dislocation structure in the thermal resistance.

\begin{table}[h!]
    \begin{center}
    \caption{Thermal boundary resistance ($R\sub{K}$; m$^2$K/GW) comparison to previous theoretical and experimental literature results for the Si-Si symmetric twist boundary}
    \begin{tabular}{|c|c|c|c|}
    \hline 
        \multicolumn{2}{|c|}{} & \textbf{Literature} $R\sub{K}$& 
        \textbf{This Study} $R\sub{K}$  \\
        \hline
        $T$ (K) & Angle $\theta$ & MD\cite{Schelling2004, Bohrer2017, Ju2013} & Born von Karman\\
        \hline
        500 & 11.42$^{\circ}$ & 0.61, 0.76, 1.1 & 0.30 \\
         \hline
         \multicolumn{2}{|c|}{} & Experimental\cite{Xu2018}& \\
         \hline
         300 & 6.9$^{\circ}$ & 9.0 & 0.21 \\
         \hline
         300 & 3.4$^{\circ}$ & 6.7 & 0.13\\
         \specialrule{.2em}{.1em}{.1em} 
         300 & $\frac{R_\mathrm{K}( 6.9^{\circ})}{R_\mathrm{K}(3.4^{\circ})}$ & 1.3 & 1.6\\
         \hline
        \end{tabular}
    \label{tab:twist_comp}
\end{center}
\end{table}

\begin{figure}
    \centering
    \includegraphics[width = 0.6\textwidth]{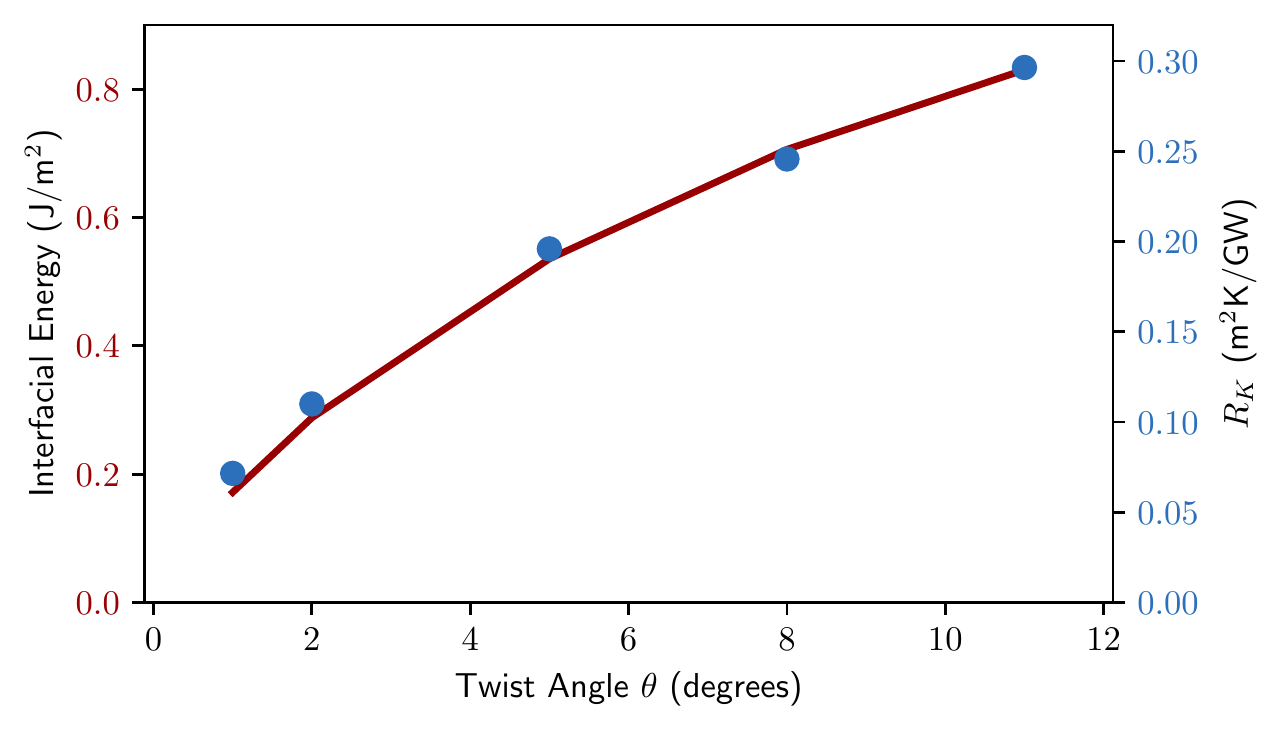}
    \caption{Thermal boundary Resistance ($R_K$) correlates with Read-Shockley grain boundary energy, as observed experimentally.}
    \label{fig:gb_energy}
\end{figure}

Finally, experimental investigations of twist boundary $R\sub{K}$ show a correlation with the Read-Shockley grain boundary energy\cite{ReadShockley}, which captures the strain energy produced by the dislocation structure at the grain boundary\cite{Xu2018, Tai2013}. This observation corroborates the idea that dislocation strain is essential to understand the origins of interfacial thermal resistance. The Read-Shockley grain boundary energy is given by:

\begin{equation}
    E = \frac{Gb}{4\pi(1 - \nu)}\theta(A - \mathrm{ln}(\theta)),
\end{equation}

\noindent with dependencies on the misorientation angle $\theta$, Burgers vector $b$, bulk modulus $G$, and Poisson ratio $\nu$. The $A$ factor captures the ratio between the dislocation core energy and strain energy contributions at the grain boundary. We set $A$ equal to 0.23, following the previous work of Tai \textit{et al.}\cite{Tai2013}, for the simple purposes of demonstrating the correlation with $R\sub{K}$. As shown in Figure \ref{fig:gb_energy}, $R\sub{K}$ from the twist boundary model closely trends with the Read-Shockley strain energy, as expected.

\section{Semicoherent Heterointerface Scattering}
\label{section:hetero}

In this section, we apply the formalism of Section \ref{subsct:scatt_theory} to a semicoherent heterointerface. 

\subsection{Misfit Dislocation Strain and Acoustic Mismatch} \label{subsct:hetint_details}
Maintaining the geometry of Section \ref{sub:scatt_kin}, the semicoherent heterointerface is defined in the yz-plane with two interpenetrating arrays of dislocations with misfit edge character\cite{Yugui1999}. As in the previous case of the tilt boundary\cite{Hanus2018}, the deformation tensor is broken down into dilatational strain ($\epsilon_\mrm{\Delta}$), shear strain ($\epsilon_\mrm{S}$), and rotation ($\epsilon_\mrm{R}$), which act as independent scattering sources. These strain fields are related to those of the tilt boundary by a simple rotation, which places the extra half plane along the $x$-axis, perpendicular to the boundary. They are given in real space in Section \ref{supp:hetint}. Table \ref{tbl:het_strain_fields} lists the Fourier strain components for a single misfit edge dislocation in both the YZ and ZY arrays.

\begin{table}[h]
\caption{Heterointerface Fourier Strain Field Components}
\label{tbl:het_strain_fields}
\centering
\large
\newcolumntype{C}{>{\centering\arraybackslash} m{6cm} } 
\begin{tabular}{|C|C|}
\hline
   YZ array  & ZY array \\
\hline
\vspace{5pt}
    $\mathlarger{\widetilde\epsilon_{\Delta} = \frac{ib(1 - 2\nu)}{(1-\nu)}\frac{Q_x}{(Q_x^2 + Q_y^2)}}$ &
 \vspace{5pt}
    $\mathlarger{\widetilde\epsilon_{\Delta} = \frac{ib(1 - 2\nu)}{(1-\nu)}\frac{Q_x}{(Q_x^2 + Q_z^2)}}$\\[20 pt] 
 \hline 
 \vspace{5pt}
    $\mathlarger{\widetilde\epsilon_{S} = \frac{-ib}{(1-\nu)}\frac{Q_yQ_x^2}{(Q_x^2 + Q_y^2)^2}}$ &
 \vspace{5pt}
 $\mathlarger{\widetilde\epsilon_{S} = \frac{-ib}{(1-\nu)}\frac{Q_zQ_x^2}{(Q_x^2 + Q_z^2)^2}}$\\[20pt]
\hline
 \vspace{5pt}
 $\mathlarger{\widetilde{\epsilon}_{R}=\frac{-2ibQ_{y}}{(Q_{x}^{2}+Q_{y}^{2})}}$ &
 \vspace{5pt}
 $\mathlarger{\widetilde{\epsilon}_{R}=\frac{-2ibQ_{z}}{(Q_{x}^{2}+Q_{z}^{2})}}$\\[20pt]
 \hline
\end{tabular}
\end{table}


Figure \ref{fig:misfit} shows a normal component of the strain field from a single misfit dislocation array, and the cross-section reveals a step function change in the dilatation at the interface\cite[pp.695-697]{CaiNix}. In fact, setting $Q_y = 0$ in the YZ array or $Q_z = 0$ in the $zy$ array, yields $\widetilde{\epsilon}_{\Delta} \propto i/Q_x$, which is precisely the Fourier transform of the Heaviside step function. As explained in Section \ref{sct:intro}, this long range dilatational strain effect is artificial, since the reference lattice parameter differs on either side of the interface. Therefore, the dilatational strain at the long-wavelength limit is subtracted and treated via the acoustic impedance mismatch term described by Equation \ref{eqn:V_tilde_R}.

Recently, Varnavides \textit{et al.} introduced the strain mismatch model (SMM) \cite{Varnavides2019}, providing an \textit{ab initio} framework for inelastic phonon scattering due to an interfacial strain perturbation. The SMM method is applied to treat a similar physical system, studying the dilatational strain scattering from a misfit dislocation array. By following the treatment of Carruthers (Eq. 4.91 of Ref. \cite{Carruthers1961}), the derived scattering rate is found to be independent of the dislocation spacing, and as far as we can interpret, neglects the periodic strain fields local to the interface, which we find to be important in this work. Both previous works \cite{Varnavides2019, Carruthers1961} additionally treat the step-change in dilatation at a misfit dislocation array as a source of anharmonic strain scattering, which differs from the acoustic mismatch approach taken here. 

\subsection{Heterointerface Results and Discussion}

\begin{figure}[h!]
    \centering
    \includegraphics[width = \textwidth]{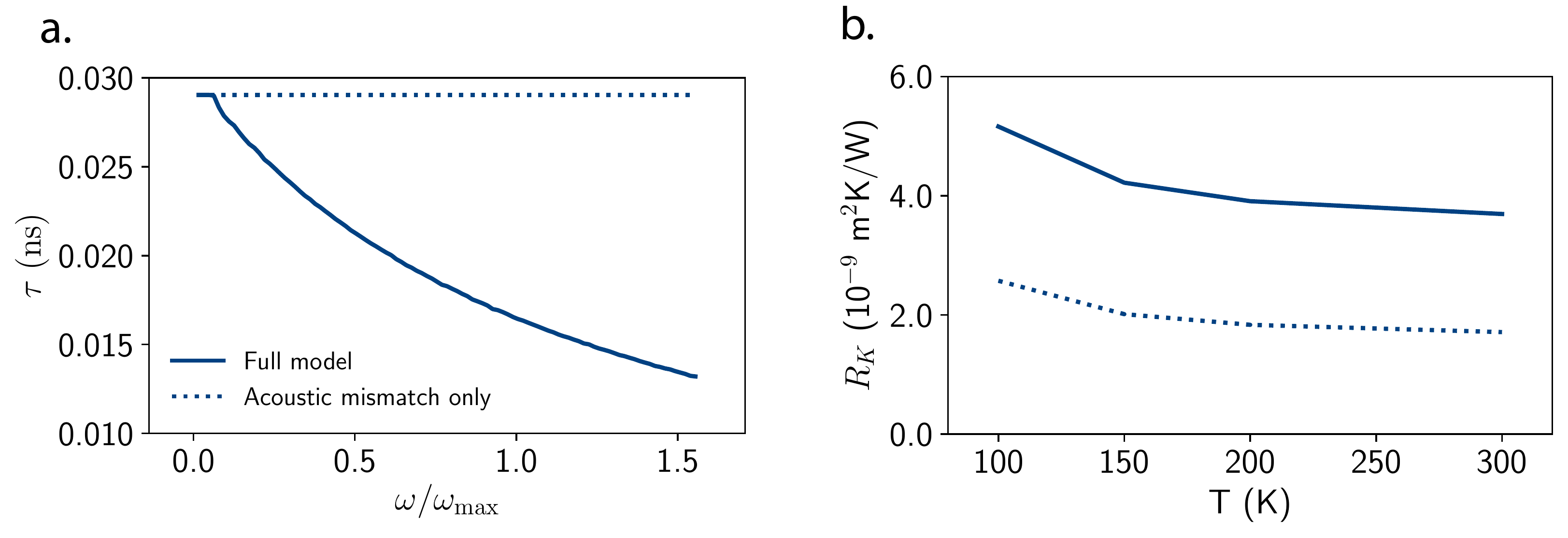}
    \caption{Si-Ge heterointerface scattering using a Born-von Karman phonon dispersion (a) Spectral relaxation time for a Si-Ge heterointerface with a misfit dislocation spacing of 7 nm, comparing model with acoustic mismatch and dislocation strain (solid) to the acoustic mismatch effect alone (dotted) (b) Thermal boundary resistance predictions versus temperature from the heterointerface model with (solid) and without (dotted) dislocation strain scattering}
    \label{fig:hetint_plots}
\end{figure}

The calculations below are performed for a Si-Ge interface using the parameters found in the Supplementary Table \ref{tab:param_vals}. As in the tilt and twist boundary examples, the heterointerface relaxation time crosses over between planar-defect and linear-defect scattering. However, as expected due to the larger acoustic mismatch, the thermal resistance is significantly larger than in the twist boundary case. The acoustic mismatch effect alone, however, contributes only about 50\% of the full thermal boundary resistance (see Figure \ref{fig:hetint_plots}) predicted by the model, indicating a significant contribution of misfit dislocation strain scattering to the thermal resistance.


Table \ref{tab:hetint_comp} shows the thermal boundary resistance ($R\sub{K}$) results of our method  assuming a Born-von Karman (BvK) model for the phonon dispersion. The results show good agreement with previous calculations on Si-Ge heterojunctions using the diffuse mismatch model (DMM) and molecular dynamics (MD) simulations\cite{Swartz1989}. In the diffuse mismatch model, the overlap in the phonon density of states on either side of the interface determines the transmission probability. In contrast, the molecular dynamics simulation uses no model or assumption about the phonon scattering mechanism. We compare to the results from the largest simulation cell trialled in each of the previous MD studies\cite{Landry2009, Zhan2015}. Our results are also in line with the atomistic Green's function (AGF) approach, which circumvents the lattice dynamical matrix equation, and instead studies the impulse response of the system\cite{Li2012}. The AGF work additionally studied the influence of alloying at the interface and observed that $R\sub{K}$ doubles with an alloy layer of just 1 nm.
Finally, in the scattering mismatch model (SMM)\cite{Varnavides2019}, phonon transmission coefficients are evaluated through an iterative solution of the phonon Boltzmann transport equation to predict $R\sub{K}$. The SMM predicts a larger $R\sub{K}$, likely due to the first-principles anharmonicity treatment and differences in the treatment of the dilatation at the interface (see Section \ref{subsct:hetint_details}). 

The thermal boundary resistance $R\sub{K}$ of the Si-Ge interface has also been investigated experimentally. The through-film thermal conductivity of superlattice films can be converted to a value for $R\sub{K}$ by assuming that bulk phonon scattering is negligible\cite{Varnavides2019}, and the results for the largest reported period $L$ (i.e. thickness of interlayers) are summarized in Table \ref{tab:hetint_comp}. While the $R\sub{K}$ calculated here is comparable to the superlattice measurements, it is important to note that its value is affected by the coherent phonon dynamics present in superlattices. Additionally, Wang \textit{et al.}\cite{Wang2020} measured the $R\sub{K}$ of a Si film bonded to a Ge substrate using the 3\,$\omega$ method, and report an order of magnitude larger thermal boundary resistance. In their study, however, a $\sim$3 nm alloy layer is shown to form with additional interdiffusion persisting for $\sim$10 nm around the interface, and this alloying effect predominates the interfacial thermal resistance. Interface quality can therefore have order of magnitude effects on the thermal transport\cite{Hopkins2013}.

\begin{table}[h!]
    \centering
\caption{Room temperature thermal boundary resistance ($R\sub{K}$; m$^2$K/GW) with comparison to theoretical and experimental literature results for Si-Ge heterointerface}
    \begin{tabular}{|p{3 cm}|p{2 cm}|p{2 cm}|p{2 cm}|p{2 cm}|p{2 cm}|p{2 cm}|}
    \hline
    \multicolumn{7}{|c|}{\textbf{Computational or Theoretical}}\\
    \hline 
        \textbf{This work} & 
        \multicolumn{2}{|c|}{Ref. \cite{Landry2009}} & \multicolumn{2}{c|}{Ref. \cite{Zhan2015}}&
        Ref. \cite{Varnavides2019}& Ref. \cite{Li2012}\\
        \hline
         \textbf{BvK} & MD& DMM & MD & DMM & SMM & AGF\\
        \hline
         \textbf{3.75} & 2.83 & 2.40 & 3.00 & 3.71 &  5.22 & 3.36\\
         \hline
        \end{tabular}
    \begin{tabular}{|p{3 cm}|p{3 cm}|p{3 cm}|p{3 cm}|}    
    \multicolumn{4}{|c|}{\textbf{Experimental (3\,$\omega$ method)}}\\
    \hline
    Ref. \cite{Xu2018}& Ref. \cite{Borca2000}& \multicolumn{2}{c|}{Ref. \cite{Lee1997}} \\
    \hline
    Bonded Films & Superlattice\newline (L = 14 nm) & Superlattice\newline (L = 15 nm)& Superlattice\newline (L = 27.5 nm) \\
    \hline
    31.4 & 2.14 & 3.62 & 6.28\\
    \hline
    \end{tabular}
    \label{tab:hetint_comp}
\end{table}

\section{Conclusion}
\label{section:conclude}

In this study, we have focused on the thermal resistance of special low-energy interfaces, which can be decomposed into periodic arrays of dislocations. Specifically, we provide a formalism for the scattering effects of two orthogonal dislocation arrays combining to form a cross-grid, and apply this model to describe symmetric twist boundaries and semi-coherent heterointerfaces. Because our model explicitly incorporates information about the dislocation structure, we can capture the effects of dislocation spacing and strain energy on scattering rates and transport coefficients. Additionally, whereas standard models of boundary scattering yield $\omega$-independent, planar defect scattering, this model explicitly shows the onset of $\omega$-dependent scattering stemming from the interactions of mid-to-high frequency phonons with interfacial dislocations. Finally, we address long-range deformations at the interface with an acoustic mismatch scattering treatment, within our perturbation theory framework. The transport predictions from this model agree well with results from molecular dynamics, but generally underestimate experimentally measured thermal resistances. However, this discrepancy may demonstrate the influence interfacial roughness and defect decoration can have on thermal resistance. The discussion presented here on interfacial thermal resistance, with comparisons of grain boundary type, angle, and energy, can accelerate a ICME microstructure engineering framework for thermal materials.

\section*{Code Availability}
All scripts used to implement this boundary scattering model can be accessed at: \url{https://github.com/RamyaGuru/BoundaryScattering}

\section*{Acknowledgments}
This work was performed under the following financial
assistance award 70NANB19H005 from U.S.
Department of Commerce, National Institute of
Standards and Technology as part of the Center for
Hierarchical Materials Design (CHiMaD). Samuel Graham and Riley Hanus acknowledge support from the Office of Naval Research under a MURI program (Grant No. N00014-18-1-2429) and Air Force Office of Scientific Research under a MURI program (Grant No. FA9550-18-1-0479).

\pagebreak

\beginsupplement

\section{Scattering Rate Calculation}
\label{supp:scatt_rate_details}
This section outlines the main steps to arrive at the final working formula (Eq. \ref{eq:tauinv_WorkingFormula}) for the scattering rate $\Gamma$ due to a dislocation grid. As before, the YZ array is composed of dislocation lines oriented along the $z$-direction that are periodically spaced by $D$ along the $y$-direction. Conversely, the ZY array has dislocation lines oriented along the $y$-direction that are periodically spaced by $D$ along the $z$-direction. Finally, the linear density of planar defects in the sample is given by $n_D=1/L_x$. For ease of this discussion, we will start with an alternative representation of Fermi's Golden Rule, which relates $\Gamma$ to the number density of defects ($n\sub{b}$), squared matrix element ($|M|^2$), and a phase space factor ($g$) capturing all the final phonon states $\mathbf{q'}$ that the incident phonon could scatter into:

\begin{equation}
    \label{eqn:FGR_real}
    \Gamma(\mathbf{q}) = n\sub{b}|M|^2g(\omega_{\mathbf{q}}).
\end{equation}

\noindent Here, $n\sub{b}$ is the linear density of boundaries in the material and is approximately equal to $1/L$, where $L$ is the average grain size in the material. As shown in Section \ref{sub:scatt_kin}, the scattering rate from the YZ and ZY dislocation grid can be computed separately. Therefore, here we will focus on deriving the scattering rate from the YZ array ($\Gamma\sub{YZ}$), and the $\Gamma\sub{ZY}$ term can be written by analogy by enforcing periodicity along the $z$-direction.

Starting with Equation \ref{eqn:Vtilde_sum} (reproduced below), the Fourier transform of the periodic scattering potential for the $\mrm{YZ}$ array is:

\begin{align}
\widetilde{V}\sub{YZ} & =\iint dx\,dy\sum_{n=-\infty}^{\infty}V_{1}(x,y-nD)e^{-i(Q_{x}x+Q_{y}y)}\nonumber \\
 & =\sum_{n=-\infty}^{\infty}e^{-iQ_{y}nD}\widetilde{V}_{1}(Q_{x},Q_{y}).
\end{align}

\noindent Here, $V_1$ is the real-space strain field around a single dislocation defect. We apply the Poisson summation formula to explicitly show the Dirac comb:
\begin{equation}
\sum_{n=-\infty}^{\infty}e^{-iQ_{y}nD}=\frac{2\pi}{D}\sum_{n'=-\infty}^{\infty}\delta(Q_{y}-Q_{n'}),\qquad\Bigl(Q_{n'}=\frac{2\pi n'}{D}\Bigr).\label{eq:PoissonSum}
\end{equation}

\noindent Hence, $\widetilde{V}\sub{YZ}$ can be written as:
\begin{align}
\widetilde{V}\sub{YZ}&=\frac{2\pi}{D}\sum_{n'=-\infty}^{\infty}\delta(Q_{y}- Q_{n'})\widetilde{V}_{1}(Q_{x},Q_{y})\\
 &=\frac{2\pi}{D}\sum_{n'=-\infty}^{\infty}\delta(Q_{y}- Q_{n'})\widetilde{V}_{1}(Q_{x},Q_{n'}).
\label{eq:Vn-squig-sqrd}
\end{align}

We then square this result and integrate over all values $\mathbf{q'}$ when computing the squared matrix element. Invoking Fermi's Golden Rule (Equation \ref{eqn:FGR}), the scattering rate for a phonon state $\mathbf{q}$ is then,
\begin{equation}
\Gamma_\mrm{YZ}(\mathbf{q})=\frac{n\sub{b}}{\hbar^{2}D^{2}}\sum_{\widetilde{n}=-\infty}^{\infty}\iiint d^{3}\mathbf{q}^{\prime}\,\delta(Q_{z})\delta(\omega_{\mathbf{q}}-\omega_{\mathbf{q^{\prime}}})\left|\widetilde{V}\sub{YZ}(Q_{x},Q_{n'})\right|^{2}(1-\hat{\mathbf{q}}\cdot\hat{\mathbf{q}}').\label{eq:Gamma-1_simplified_final}
\end{equation}

\noindent This is the same as Equation S41 of Hanus \textit{et al.}\cite{Hanus2018}. The phase space term is implicitly enforced by the momentum and energy conservation laws imposed by the $\delta$-functions. These conservation rules restrict
this integral to a discrete set of available $\mathbf{q}^{\prime}$
states. Equations S42 - S48 of Hanus \textit{et al.}\cite{Hanus2018} re-express these $\delta$-functions explicitly in terms of $\mathbf{q'}$ to arrive at the final expression for $\Gamma$:

\begin{equation}
\Gamma_\mrm{YZ}(\mathbf{q})=\frac{n\sub{D}}{\hbar^{2}v_{\mathrm{g}}D^{2}}\sum_{n'=-\infty}^{\infty}\sum_{\sigma=\pm}\frac{(q_{x}^{2}\mp q_{x}(q_{x}^{2}+2q_{y}Q_{n'}-Q_{n'}^{2})^{1/2}+q_{y}Q_{n'})}{q(q_{x}^{2}+2q_{y}Q_{n'}-Q_{n'}^{2})^{1/2}}\left|\widetilde{V}_{1}(Q_{x,n'\sigma},Q_{n'})\right|^{2}.\label{eq:tauinv_WorkingFormula}
\end{equation}
This is the working formula that we use for numerical calculation of scattering rate due to the $\mrm{YZ}$-array of dislocations, and equivalently for the $\mrm{ZY}$-array by switching the $y$ and the $z$ components. 

Finally, to get the spectral relaxation time, we must average over the incident phonon wavevector direction $\hat{q}$. We compute $\tau(\omega)$ as the weighted orientational average of the inverse scattering rate, 

\begin{equation}
    \tau(\omega) = \frac{\iint \Gamma^{-1} q_x^2 d\Omega}{\iint q_x^2 d\Omega} = \frac{3}{4\pi} \iint \Gamma^{-1} \frac{q_x^2}{q^2} d\Omega,
    \label{eq:tau_spectral}
\end{equation}

\noindent where $d\Omega$ is the element of solid angle for $\hat{q}$.

The spectral relaxation time can then be used to compute transport properties such as the lattice thermal conductivity in the phonon gas model,

\begin{equation}
    \kappa\sub{L} = \frac{1}{3}\int C_s(\omega) v\sub{g}^2 \tau(\omega) d \omega,
\end{equation}

\noindent using the expression,

\begin{equation}
    C_s(\omega) = \frac{3\hbar^2}{2\pi^2k\sub{B}T^2}\frac{\omega^4e^{\hbar\omega/k\sub{B}T}}{v\sub{g}v\sub{p}^2(e^{\hbar\omega/k\sub{B}T} - 1)^2}
\end{equation}
for the spectral heat capacity $C_s(\omega)$.

\section{Relating Landauer and interfacial scattering theory \label{sec:Landauer-pert}}

\begin{figure}[t]
    \begin{center}
    \includegraphics[scale=0.6]{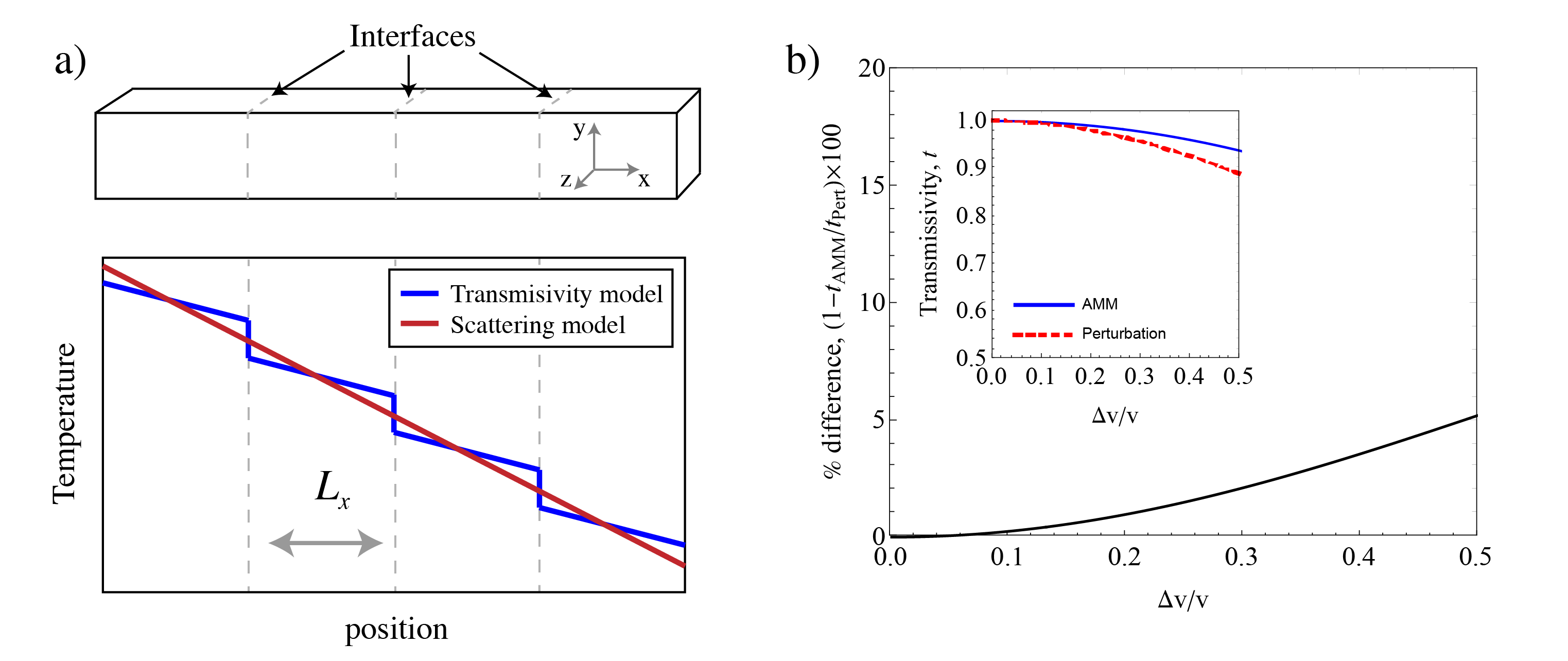} 
    \end{center}
    \caption{a) A schematic illustration of two common models used to describe heat conduction in materials with interfaces. The blue line depicts the Landauer based model where a thermal boundary resistance arising from the conduction channel having a interfacial transmission probability or $t(\omega)>0$,  induces a sharply localized drop in temperature. The red line depicts a model based on phonon scattering theory and Matthiessen's rule, where each scattering mechanism contributes a scattering rate  ($\tau(\omega)^{-1}$), and together modify the materials thermal conductivity homogeneously. b) A comparison between the transmissivity calculated using classical acoustic mismatch (AMM) theory and quantum perturbation theory. The two cases differ by no more than 5\% up to $\Delta v/v=0.5$. \label{fig:Mattheissen-vs-Landauer}}
\end{figure}

There are several ways to model thermal transport in systems containing interfaces. Two common approaches are the Landauer formalism, which defines a phonon transmissivity and thermal boundary resistance, and scattering theory, which models the interfacial thermal resistance via an additional relaxation time that modifies the  thermal conductivity of the total system (bulk material plus interfaces). Figure \ref{fig:Mattheissen-vs-Landauer}a depicts the temperature profiles implied by the Landauer theory (blue line) versus scattering theory (red line) approaches. If the mean free path of a phonon is equal to or longer than $L_x$, the homogeneously sloped red line better represents reality. If the mean free path is much shorter than $L_x$, the distance between interfaces, inhomogeneities in the temperature gradient are expected near interfaces, and the blue line may be more appropriate. It is important to keep in mind that materials tend to have a wide distribution of mean free paths, so the red line may better describe low-frequency phonons, while the blue line may better describe high-frequency phonons.  Regardless, we can mathematically relate these two frameworks and define a relationship between the relaxation time and transmissivity. 

The spectral thermal conductivity of the bulk, interface-free system is given by 
\begin{equation}
\kappa_\mathrm{bulk}(\omega)=\frac{1}{3}C_\mrm{s}(\omega)v_\mathrm{g}(\omega)^2\tau_\mathrm{bulk}(\omega),
\end{equation}
where $\tau_\mrm{bulk}$ describes scattering in the bulk material and may contain phonon-phonon and phonon-point defect scattering, for example. The thermal resistance due to a slab of this bulk material of length of $L_x$ is $(L_x \kappa_\mrm{bulk})^{-1}$.

\subsection{Landauer theory}
Within the Landauer approach, we define a spectral thermal boundary conductance $h_\mrm{B}$ based on the transmissivity from side 1 to side 2 of the interface $\alpha_{12}$ and the reverse direction $\alpha_{21}$ as: 

\begin{equation}
    h_\mrm{B}(\omega) = \frac{1}{4} C_\mrm{s}(\omega)v_\mrm{g}(\omega)\left( \frac{\alpha_{12}(\omega)}{1 - \overline{\alpha}(\omega)} \right),
\end{equation}
where $\overline{\alpha} (\omega) = (\alpha_{12}(\omega) + \alpha_{21}(\omega))/2$. The thermal boundary resistance of the interface is then defined by
\begin{equation}\label{eqn:Rk_SI}
1/R\sub{K} =h_\mrm{B}= \int_0^{\omega\sub{m}} h_\mrm{B}(\omega) d\omega.
\end{equation}
Since the bulk material and interfaces are in series in this model, the total spectral thermal resistance is given by the sum of the bulk and interface resistances:
\begin{equation}
(L_x\kappa_\mrm{L}(\omega))^{-1}=(L_x\kappa_\mrm{bulk}(\omega))^{-1}+( h_\mrm{B}(\omega))^{-1}.
\label{eq:kappa_int_Land}
\end{equation}

\subsection{Interfacial scattering}\label{suppsct:boundary_scatt}
The second interfacial resistance method treats the interface as an additional scattering mechanism, which can reduce the total phonon lifetime. The total thermal resistance of the same length of material $L_x$ is given by Matthiessen's rule as the sum of the resistance due to bulk phonon scattering and the resistance due to boundary scattering, 
\begin{equation}
(L_x\kappa_\mrm{L}(\omega))^{-1}=(L_x\kappa_\mrm{bulk}(\omega))^{-1}+(L_x\kappa_\mrm{B}(\omega))^{-1}. \label{eq:kappa_int_Pert}
\end{equation}
In this work, we refer to the interface relaxation time as $\tau\sub{B}$, giving
\begin{equation}
    \kappa_\mrm{B}=\frac{1}{3}C_\mrm{s}(\omega)v_\mrm{g}(\omega)^2\tau\sub{B}(\omega).
    \label{eq:kappa_B}
\end{equation}
We can rewrite Eqs. \ref{eq:kappa_int_Pert} and \ref{eq:kappa_B} as
\begin{equation}
\kappa_\mrm{L}(\omega)=\frac{1}{3}C_\mrm{s}(\omega)v_\mrm{g}(\omega)^2\tau_\mrm{tot}(\omega),
\end{equation}
with $\tau_\mrm{tot}(\omega)$ given by the sum of scattering rates due to bulk processes such as phonon-phonon scattering ($\tau\sub{ph}$) and interface scattering ($\tau\sub{B}$):
\begin{equation} 
\tau_\mrm{tot}(\omega)^{-1}=\tau_\mrm{ph}(\omega)^{-1}+\tau\sub{B}(\omega)^{-1}.
\end{equation}

By equating Equations \ref{eq:kappa_int_Land} and \ref{eq:kappa_int_Pert}, we get $L_x\kappa_\mrm{B}(\omega)=h_\mrm{B}$. Now, when the relation $\tau_{12}=\tau_{21}(v_{\mrm{g},1}/v_{\mrm{g},2})$ is obeyed---which we will prove to hold true for our acoustic mismatch scattering potential next---we obtain Equation \ref{eq:alpha12_to_tau}. Here, the 12 subscript explicitly indicates that the phonon is approaching the interface from material 1, and vice versa for 21. This notation is omitted in the main text. 

We now aim to directly compare the treatment of acoustic mismatch via our perturbation theory approach and the classical AMM treatment. To do so, we define an interface between two Debye solids where $v_{1}=v+\Delta v/2$, $v_{2}=v-\Delta v/2$, and density is unchanged, $\rho_1=\rho_2$. Following the procedure presented by Hanus, Garg, and Snyder \cite{Hanus2018}, and calculating the relaxation time due to the scattering potential given in Eq. \ref{eq:V_AM}, we find
\begin{equation}
    \tau_{12}(\mathbf{q})^{-1}=\frac{\Delta v^2 }{2 L_z v_{1}|\sin \theta \cos \phi|}.
    \label{eq:tau_AMM}
\end{equation}
At the acoustic mismatch limit, the planar defect produces a specular reflection, which scatters back into the density of states of side 1. The factor of $v\sub{g}$ in the denominator, which comes from the density of states contribution, is therefore equal to $v_1$. The spectral relaxation time can be obtained by using Eq. \ref{eq:tau_AMM} and Eq. \ref{eq:tau_spectral}:

\begin{equation}
    \tau_{12}(\omega)=\frac{3}{2}\frac{L_x v_\mrm{1}}{\Delta v^2}.
    \label{eq:tau12}
\end{equation}

\noindent Note that the relation $\tau_{12}=\tau_{21}(v_{2}/v_{1})$ holds, since the factor of $v_\mrm{1}$ is associated with the density of states into which the phonon is scattering (i.e. reflection back into material 1). The details regarding the matrix element contribution to $\tau$ result in no change to the final result when inverting the problem from 12 to 21. 

Using Eq. \ref{eq:tau12} with Eq. \ref{eq:alpha12_to_tau}, we obtain for the perturbation theory transmissivity,
\begin{equation}
    t_\mrm{Pert}=\frac{1}{\frac{1}{2} \left( \frac{\Delta v}{v} \right)^2+1}.
\end{equation}
In the classical AMM theory, the transmissivity is given by \cite{Swartz1989}
\begin{equation}
t_\mrm{AMM}=\frac{4 Z_1 Z_2}{(Z_1 + Z_2)^2},
\label{eq:t_AMM}
\end{equation}
where $Z_i=\rho_i v_{\mrm{s},i}$ is the acoustic impedance of side $i$, and $\rho_i$ is its mass density. The two cases are directly compared in Figure \ref{fig:Mattheissen-vs-Landauer}b and they differ by no more than 5\% up to $|\Delta v/v|=0.5$.

\section{Twist Boundary Strain Field Details}
\label{suppsct:twist}
The displacement vector field for a screw dislocation has only one non-zero component oriented along the line of the dislocation. Therefore, a dislocation with Burgers vector ($b$) parallel to $\hat{z}$ has the following displacement field\cite[p. 60]{Loethe1982}:

\begin{align}
\mathbf{u}_{n} & =(0,0,u_{z}),\label{eq:screw-disp-z}\\
u_{3,n} & =\frac{b}{2\pi}\arctan\left(\frac{y}{x}\right).\nonumber 
\end{align}

\noindent The displacement produces a pure shear state, with only two non-zero strain components:
\begin{align}
\epsilon_{xz}=\epsilon_{zx}&=\frac{1}{2}\frac{\partial u_{z}}{\partial x}=-\frac{by}{4\pi(x^{2}+y^{2})},\\
\epsilon_{yz}=\epsilon_{zy}&=\frac{1}{2}\frac{\partial u_{z}}{\partial y}=\frac{bx}{4\pi(x^{2}+y^{2})}.\nonumber
\end{align}

Next, we will consider the $\mrm{YZ}$ array of screw dislocations spaced by $D$ along the $y$-axis in order to model a low-angle twist boundary. The stress components from the dislocation array $\epsilon_{ij}^{\mrm{YZ}}$ can be determined from the following summation:

\begin{equation}
    \epsilon_{ij}^{\mrm{YZ}} = \sum_{n=-\infty}^{\infty}\epsilon_{ij}(x,y-nD).
\end{equation}

\noindent Analytic solutions to the above summation can be obtained, and are shown below \cite[pp. 698-700]{CaiNix}:

\begin{align}\label{eqn:real_spc_twist}
    \epsilon_{xz}^{\mrm{YZ}} = -\frac{b}{2D}\left(\frac{\sin{(2\pi y/D)}}{\cosh{(2\pi x/D) - \cos{(2\pi y/D)}}}\right),\\
    \epsilon_{yz}^{\mrm{YZ}} = \frac{b}{2D}\left(\frac{\sinh{(2\pi x/D)}}{\cosh{(2\pi x/D) - \cos{(2\pi y/D)}}}\right)\nonumber.
\end{align}
One can then evaluate the limit as $|x|\to\infty$:
\begin{align}
    \lim_{x\to\infty}\epsilon_{xz}^{\mathrm{YZ}} &= 0,\\
     \lim_{x\to\infty}\epsilon_{yz}^{\mathrm{YZ}} &= \mathrm{sgn}(x)\frac{b}{2D}.\nonumber
\end{align}
The $\epsilon_{yz}^{\mathrm{YZ}}$ shear strain component persists at the long-range limit, converging to a constant value. This is energetically prohibitive for the twist boundary as a whole and shows the importance of including the $\mrm{ZY}$ array of dislocations with sense vector along the $y$-axis, periodically spaced on the $z$-axis. The strain components from this array are the negative of Equation \ref{eqn:real_spc_twist} with $y$ and $z$ swapped, and their long-range limits are,

\begin{align}
    \lim_{x\to\infty}\epsilon_{xz} &= 0,\\
     \lim_{x\to\infty}\epsilon_{yz} &= -\mathrm{sgn}(x)\frac{b}{2D},
\end{align}

\noindent which exactly cancel the far-field strain of the first array if both share the same $b/D$ ratio\cite{CaiNix}.

\section{Heterointerface Strain Field Details}
\label{supp:hetint}
As with the twist boundary, the heterointerface is taken to lie in the $yz$-plane with two interpenetrating arrays of dislocations, but now with edge character. Therefore, the strain fields are essentially equivalent to the tilt boundary case, requiring only a rotation such that the extra half-plane points in the $x$ direction. They are given below for one dislocation through the origin in each of the two arrays. The notation follows Ref. \cite{Hanus2018}.



\begin{table}[h]
\caption{Heterointerface Strain Field Components}
\label{tbl:het_strain_fields_real}
\centering
\large
\newcolumntype{C}{>{\centering\arraybackslash} m{6cm} } 
\begin{tabular}{|C|C|}
\hline
   YZ array  & ZY array \\
\hline
\vspace{5pt}
    $\mathlarger{\epsilon_{\Delta} = \frac{-b(1 - 2\nu)}{2\pi(1-\nu)}\frac{x}{(x^2 + y^2)}}$ &
 \vspace{5pt}
    $\mathlarger{\epsilon_{\Delta} = \frac{-b(1 - 2\nu)}{2\pi(1-\nu)}\frac{x}{(x^2 + z^2)}}$\\[20 pt] 
 \hline 
 \vspace{5pt}
    $\mathlarger{\epsilon_{S} = \frac{b}{4\pi(1 - \nu)}\frac{y(y^2 - x^2)}{(x^2 + y^2)^2}}$ &
 \vspace{5pt}
 $\mathlarger{\epsilon_{S} = \frac{b}{4\pi(1 - \nu)}\frac{z(z^2 - x^2)}{(x^2 + z^2)^2}}$\\[20pt]
\hline
 \vspace{5pt}
 $\mathlarger{\epsilon_{R} = \frac{b}{\pi}\frac{y}{x^2 + y^2}}$ &
 \vspace{5pt}
 $\mathlarger{\epsilon_{R} = \frac{b}{\pi}\frac{z}{x^2 + z^2}}$\\[20pt]
 \hline
\end{tabular}
\end{table}

Again, considering the YZ array, analytic solutions exist for the real-space sum over the misfit edge dislocations periodically-spaced by $D$. The analytic solutions for the three, independent non-zero strain components $\epsilon_{ij}^{\mathrm{YZ}}$ in a Cartesian basis are\cite[pp. 695-697]{CaiNix}\cite{VanDerMerwe1950}:


\begin{align}
    \epsilon_{xx}^{\mathrm{YZ}} &= \frac{b}{4(1-\nu)D}\left[ \frac{-2\nu S_X(C_X - c_Y) + 2\pi X(C_Xc_Y - 1)}{(C_X - c_Y)^2}\right],\\
    \epsilon_{yy}^{\mathrm{YZ}} &= \frac{b}{4(1-\nu)D}\left[ \frac{2(1-\nu)S_X(C_X - c_Y) - 2\pi X(C_Xc_Y - 1)}{(C_X - c_Y)^2}\right],\\
    \epsilon_{xy}^{\mathrm{YZ}} &= \frac{b}{2(1-\nu)D}\left[ s_Y(\frac{2\pi X S_X - C_X + c_Y}{(C_X - c_Y)^2})\right],
\end{align}

\noindent where $X\equiv x/D$, $Y\equiv y/D$, $s_Y \equiv \mathrm{sin}2\pi Y$, $c_Y \equiv \mathrm{cos}2\pi Y$, $S_X\equiv \mathrm{sinh}2\pi X$, and $C_X\equiv \mathrm{cosh}2\pi X$. 

We can again evaluate the limit as $|x|\to\infty$:

\begin{align}
     \lim_{x\to\infty}\epsilon_{xx}^{\mathrm{YZ}} &= \mathrm{sgn}(x)\frac{-b\nu}{2(1-\nu)D},\\
     \lim_{x\to\infty}\epsilon_{yy}^{\mathrm{YZ}} &= \mathrm{sgn}(x)\frac{b}{D},\\
     \lim_{x\to\infty}\epsilon_{xz}^{\mathrm{YZ}} &=0. 
\end{align}

Here, the dilatation strain components ($\epsilon_{xx}^{\mathrm{YZ}}$ and $\epsilon_{yy}^{\mathrm{YZ}}$) persist in the far field limit, while the shear strain decays. Additionally, the far-field dilatational strain is not cancelled out by the ZY array. However, as noted in the text, the nonzero dilatation in the far field is artificial since the reference lattices are different on either side of the interface. We reiterate that this far-field dilatational strain is subtracted and treated with an acoustic mismatch term capturing the step function change in stiffness matrix at the interface.

\section{Christoffel Equation}\label{supp:christ_eq}
The Christoffel matrix $C$ is obtained from the rank 4 stiffness tensor $c_{ijkl}$ for a unit vector $\hat{n}$ denoting the phonon direction of propagation as follows\cite{Jaeken2016}:

\begin{equation}
    C_{ij} = \sum_{jk}n_jc_{ijkl}n_k.
\end{equation}
From this, one can then evaluate the following eigenvalue problem to arrive at the phase velocity $v_\mrm{p}$ of an acoustic phonon travelling in the direction $\hat{n}$ with polarization vector $\hat{s}$:

\begin{equation}
    \sum_{ij} \left(C_{ij} -\delta_{ij}v_\mrm{p}^2 \right)s_{j}= 0.
\end{equation}

\begin{figure}[h!]
    \centering
    \includegraphics{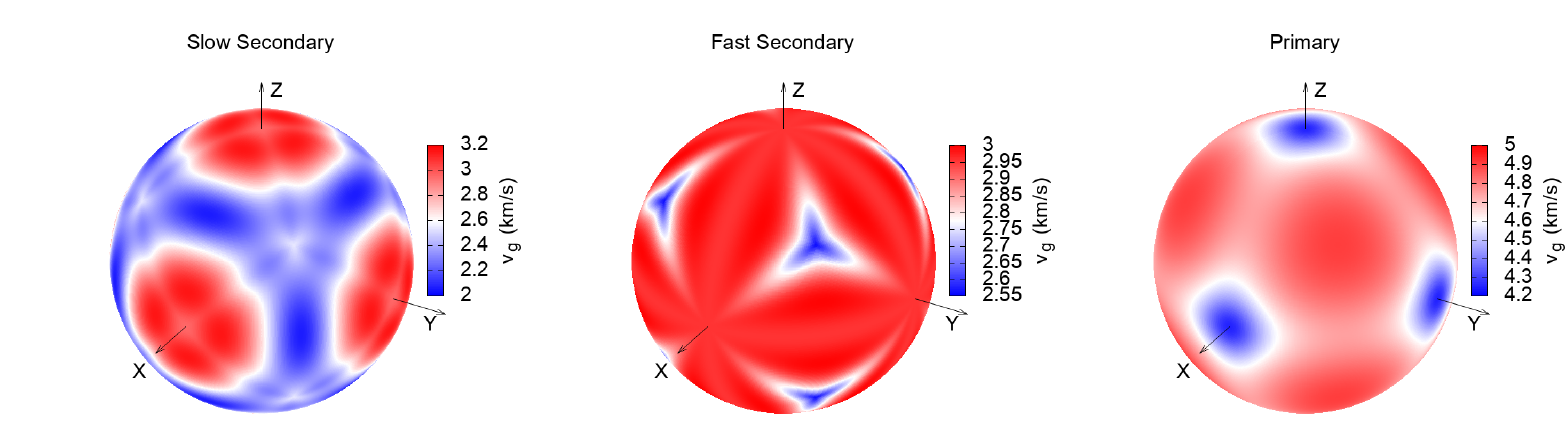}
    \caption{Group velocity slowness plots for Silicon: direction dependence of the acoustic phonon group velocities potted on a unit sphere. Fast/slow secondary correspond approximately to the two transverse branches, and primary is approximately the longitudinal branch. Produced using the \texttt{christoffel} package\cite{Jaeken2016}.}
    \label{fig:slowness_plot}
\end{figure}

\noindent By solving this equation for different $\hat{n}$, it is possible to generate a slowness surface, or diagram of the direction-dependent group ($v\sub{g}$) or phase velocity ($v\sub{p}$) of the acoustic phonons in a material. As discussed in Jaeken \textit{et al.}\cite{Jaeken2016}, the acoustic $v\sub{g}$ and $v\sub{p}$ differ slightly in terms of direction alone, as described by the power flow angle $\Psi$, where $v\sub{p} = v\sub{g}\mathrm{cos}\Psi$. The group velocity direction indicates the direction in which energy travels, which can deviate from the wavefront propagation direction described by the phase velocity. Figure \ref{fig:slowness_plot} shows the group velocity slowness plots of the three phonon polarizations for Si.

Both the twist boundary and semicoherent interface models require solving the Christoffel equations to calculate the magnitude change of phonon velocity at the boundary $\Delta v$ for a fixed incident phonon direction. In the twist boundary, $\Delta v$ comes solely from the misorientation, or equivalently, the rotation of the slowness plots at the boundary. In contrast, in the \ch{Si}-\ch{Ge} heterointerface, for example, we calculate slowness plots for both the \ch{Si} and \ch{Ge} lattices, and compute $\Delta v$ from the differences in acoustic velocity between the two materials for a fixed phonon direction. A misorientation could also be incorporated in the heterointerface by applying a relative rotation to the \ch{Ge} slowness plot with respect \ch{Si}, for example.  

\section{Model Parameters}

The model parameters are summarized in Table \ref{tab:param_vals} with the values used for the Si-Si twist boundary and Si-Ge heterointerface examples discussed in the text. Across a temperature range of 100- 800$^{\circ}$C, the temperature dependence of the input parameters due to lattice thermal expansion had a negligible impact on the phonon relaxation time predictions. However, temperature dependent inputs could be determined from quasi-harmonic DFT calculations, for example. 

\begin{table}[h!]
    \centering
        \caption{Parameters used in Model}
    \begin{tabular}{|c|c|c|}
    \hline 
       \textbf{Properties} & Silicon & Germanium\\
    \hline
       Speed of sound ($v_s$; m/s)\cite{Toberer2011}  & 6084 & 5400 \\
       Atoms per unit cell ($N$) & 2 & 2 \\
       Volume per atom ($V$; \AA$^3$)\cite{Toberer2011}& 19.7 & 22.7 \\
       Density ($\rho$, kg/m$^3$)& 2330 & 5323 \\
       Stiffness coefficients  ($c_{11}, c_{12}, c_{44}$; GPa)\cite{Hopcroft2010, Escalante2018}& 165.6, 63.9, 79.5& 126.0, 44.0, 67.7 \\
       Bulk Modulus ($G$; GPa)\cite{Hopcroft2010} & 97.83 & \\
       Gr\"uneisen parameter ($\gamma$)\cite{Hanus2018} & 1 & \\
    \hline 
    \end{tabular}
    \label{tab:param_vals}
\end{table}

The most computationally demanding portion of the model is the calculation of the spectral relaxation time, owing to the integrals over incident phonon direction and phonon frequency. Running serially on a laptop, the calculation of each spectral relaxation time value $\tau(\omega)$ takes 2.16 minutes. We find that a spline of 50 spectral relaxation time $\tau(\omega)$ values are sufficient to converge the thermal boundary resistance $R\sub{K}$. Therefore, running serially on a laptop, each thermal boundary resistance calculation takes approximately 2 hours. 


\bibliographystyle{apsrev4-2}
\bibliography{TheGrid}

\end{document}